\documentclass[journal,twocolumn, 9 pt]{IEEEtran}
\usepackage{amsmath}
\usepackage[utf8]{inputenc}
\usepackage[section]{placeins}
\usepackage{multirow}
\usepackage{subfig}
\usepackage{caption}
\usepackage{mathdots}
\usepackage{txfonts}
\usepackage{amssymb}
\usepackage[utf8]{inputenc} 
\usepackage[english]{babel} 
\usepackage{wrapfig}
\usepackage{makecell}
\usepackage{lscape}
\usepackage{fixltx2e}
\usepackage{enumitem}
\newlist{steps}{enumerate}{1}
\setlist[steps, 1]{leftmargin=1.3cm, label = Step \arabic*:}
\newlist{procedure}{enumerate}{1}
\setlist[procedure, 1]{leftmargin=2cm, label = Procedure \arabic*:}

\usepackage[numbers]{natbib}
\ifCLASSINFOpdf
   \usepackage[pdftex]{graphicx}
\else
\fi
\usepackage{array}

\usepackage{stfloats}
\usepackage{url}

\begin{document}
%
\title{Hybrid Transformer Network for Different Horizons-based Enriched Wind Speed Forecasting}
%
%
%

\author{M. MADHIARASAN$^{1*}$, Member,~IEEE, ~\IEEEmembership{$^{1*}$Department of Computer Science and Engineering, Indian Institute of Technology Roorkee,}
        PARTHA~PRATIM~ROY$^{2}$, Member,~IEEE, ~\IEEEmembership{$^{2}$Department of Computer Science and Engineering, Indian Institute of Technology Roorkee}
\thanks{Dr. M. MADHIARASAN was with the Department of Computer Science and Engineering, Indian Institute of Technology Roorkee, Roorkee, Uttarakhand, India – 247667, $^{*}$Corresponding Author, e-mail: $^{1*}$mmadhiarasan89@gmail.com, mmadhiarasan.cse@sric.iitr.ac.in, $^{1*}$ Orcid ID: 0000-0003-2552-0400.}
\thanks{Prof. PARTHA~PRATIM~ROY was with the Department of Computer Science and Engineering, Indian Institute of Technology Roorkee, Roorkee, Uttarakhand, India – 247667,  e-mail:partha@cs.iitr.ac.in, Orcid ID: 0000-0002-5735-5254.}
\thanks{Manuscript received April, 2022; revised XX, 2022.}}
%
%

\markboth{Journal of \LaTeX\ Class Files,~Vol.~xx, No.~x, April~2022}%
{Shell \MakeLowercase{\textit{Madhiarasan et al.}}: Hybrid Transformer Network for Different Horizons-based Enriched Wind Speed Forecasting for IEEE Communications Society Journals}
%



\maketitle

\begin{abstract}
 Highly accurate different horizon-based wind speed forecasting facilitates a better modern power system. This paper proposed a novel astute hybrid wind speed forecasting model and applied it to different horizons. The proposed hybrid forecasting model decomposes the original wind speed data into IMFs (Intrinsic Mode Function) using Improved Complete Ensemble Empirical Mode Decomposition with Adaptive Noise (ICEEMDAN). We fed the obtained subseries from ICEEMDAN to the transformer network. Each transformer network computes the forecast subseries and then passes to the fusion phase. Get the primary wind speed forecasting from the fusion of individual transformer network forecast subseries. Estimate the residual error values and predict errors using a multilayer perceptron neural network. The forecast error is added to the primary forecast wind speed to leverage the high accuracy of wind speed forecasting. Comparative analysis with real-time Kethanur, India wind farm dataset results reveals the proposed ICEEMDAN-TNF-MLPN-RECS hybrid model's superior performance with MAE=1.7096× 10$^{-07}$, MAPE=2.8416× 10$^{-06}$, MRE=2.8416× 10$^{-08}$, MSE=5.0206×10$^{-14}$, and RMSE=2.2407×10$^{-07}$ for case study 1 and MAE=6.1565×10$^{-07}$, MAPE=9.5005×10$^{-06}$, MRE=9.5005×10$^{-08}$, MSE=8.9289×10$^{-13}$, and RMSE=9.4493×10$^{-07}$ for case study 2 enriched wind speed forecasting than state-of-the-art methods and reduces the burden on the power system engineer.
\end{abstract}

\begin{IEEEkeywords}
Empirical Mode Decomposition, Transformer Network, Fusion, Multilayer Neural Perceptron Network, Residual Error Correction, Wind Speed and Forecasting.
\end{IEEEkeywords}

%
\IEEEpeerreviewmaketitle

\section{Introduction}
\IEEEPARstart
According to the Global Wind Report 2021\footnote[1]{https://gwec.net/global-wind-report-2021}, wind power's worldwide installed capacity is 743 GW. Thus,  the globally reduced CO2 is 1.1 billion Tonnes. As per the IRENA (International Renewable Energy Agency), as of 2020, the globally installed wind power plant is 34,367 MW concerned for offshore and 698,043 MW for on-shore capacity\footnote[2]{https://www.irena.org/wind}. Globally, all countries ramp up the wind power installed capacity to minimize global warming. The contribution of wind energy systems in power systems has increased. As per the Ministry of New and Renewable Energy report, as of 31 March 2021, India posed 39.25 GW (Gigawatt) total installed wind energy capacity\footnote[3]{https://mnre.gov.in/wind/current-status/}. Across the world, among other renewable energy resources, wind energy receives more importance. A good deal of significant wind farms are established, like Jiuquan Wind power Base (China, 20GW), Jaisalmer Wind Park (India, 1600 MW), Alta Wind Energy Center (United States, 1548 MW), etc\footnote[4]{indpower.net}. India is placed in the fourth position concerning wind energy installed capacity worldwide. The modern power system faces several challenges because of the high irregular tendency of wind-based power generation due to the volatile characteristic of wind speed. The wind energy system penetrates modern power system security, reliability, effective operation, and control depending on the correct wind speed forecasting tool. 
\par General forecasting model typified as persistent models (assumption-based forecasting approaches, suitable for very short scale horizon), physical (mathematical equations and physical law-based approach practical for forecasting timescale of medium and long term, example: NWP (Numerical Weather Prediction)), statistical methods (historical dataset based approach, an ideal choice for short-term forecasting, example: ARIMA (Autoregressive Integrated Moving Average)), an artificial neural network belongs to the statistical model type because it learns from the data set and suitable for multi-time scale forecasting, example: RNN (Recurrent Neural Network), BPN (Backpropagation Network), RBFN (Radial Basis Function Network), MLPN (Multilayer Perceptron Neural Network), deep learning neural network have overcome the limitation of the shallow neural network improve the forecasting performance by deep learning, the ability to handle massive datasets, for Example: DBN (Deep Belief Network), DNN (Deep Neural Network). and hybrid model (taking advantage of various individual models and combined to improve the forecasting performance suitable for multi-time scale forecasting) \cite{madhiarasancertain}. Planning concerns for development, expansion, and maintenance received help from the long term (period ranging from a week to years ahead) wind speed forecasting. This paper designs an astute hybrid forecasting model for wind speed forecasting on different forecasting horizons. To the best of our literature knowledge, a transformer-based combinational hybrid forecasting model with ICEEMDAN and residual error correction strategy is the first time proposed in this work for wind speed forecasting application. \\
The novelty of the proposed hybrid transformer network (ICEEMDAN-TNF-MLPN-RECS) model was not devised in the literature before and performed promising to result in highly accurate forecasting of wind speed for different horizons, converging quickly in lesser iterations with the smallest performance error index. 
\par The rest of the paper is organized accordingly. Section 2 signifies the problem definition and existing literature works. Section 3 details about the data source, performance error index, technical concepts, and framework of the proposed hybrid forecasting model. Section 4 provides the experimental analysis and subsequent performance comparative investigation discussed in Section 5. Section 6 presents the performance validity of different horizons forecasting, and finally, Section 7 summarizes and concludes the paper with a future research outlook.
\section{PROBLEM DESCRIPTION AND RELATED WORK CONCERN TO WIND SPEED FORECASTING}
The assumption-based forecasting is performed with a persistent method. The core of the physical method is physical law equations, and similarly, the heart of the statistical techniques is the dataset because forecasting models learn from the dataset. Persistent method forecasting accuracy is decaying with increasing the forecasting horizon. Physical methods possess complexity related to computation that requires high power computing devices and high cost. Inherent bias is another issue of numerical weather prediction. Statistical methods encounter local minima issues, entrap into overfitting issues, and the inability to handle big data. Shallow neural networks can predict, but they have problem concerns in handling big data and memory storage for past information, learned by the series, not through parallel learning and convergence time. Slow convergence and local minima are the two major flaws of the BPNN. To fasten the convergence speed, the momentum factor is added \cite{madhiarasan2016novel}, a concern to the neural network hyperparameter choice is critical. There is no specific theoretical rule thump is available for generic \cite{madhiarasan2016novel}. 
\par Various researchers in the literature have tried a variety of wind speed forecasting models. Some studies in the literature developed a forecasting model. The persistent method \cite{franke1995investigations} is a baseline for forecasting applications. Numerical weather prediction considering local effect-based 36 hours ahead wind speed forecasting is carried out using HIRLAM (High-Resolution Limited Area Model) \cite{landberg1994short}. Short-time interval-based wind speed forecasting using ARIMA was demonstrated in \cite{palomares2009arima}. In reference, \cite{madhiarasan2016performance} evolved various artificial neural networks, multi-time scale-based wind speed forecasting. Local recurrent neural network-based wind speed and wind power forecasting in a long-term horizon (72 hours ahead) performed by \cite{barbounis2006long}. Radial basis function network-based wind speed and power forecasting model suggested in reference \cite{junli2010wind}. Reference \cite{botha2017forecasting} presented a Support Vector Regression (SVR) based short-term wind speed forecasting model. \citet{ahmadi2020long} analyzed the tree-based learning algorithm performance based on wind power forecasting for six months ahead forecasting. They stated that the XGboost method performs better than other models. Multilayer perceptron network-based wind speed forecasting and hidden neuron-based comparative analyses were performed in \cite{madhiarasan2017comparative}. \citet{araya2018lstm} carried out the wind speed forecasting concern with multi timescale horizon using LSTM. Gated Recurrent Unit (GRU) based wind speed and power forecasting for ultra-short time horizon suggested in \cite{syu2020ultra}. Recursive strategy is introduced in the radial basis function network and shows a highly accurate wind speed forecast \cite{madhiarasan2020accurate}. \citet{zhang2015predictive} performed wind speed forecasting for short-term and long-term horizons based on the PDBM (Predictive Deep Boltzmann machine). 
\par \citet{ren2014comparative} studied short-term wind speed forecasting using ANN and SVR (Support Vector Regression) in association with various variants of EMD (Empirical Mode Decomposition). Interval prediction concern for wind speed prediction using Variational Mode Decomposition (VMD) associated particle swarm optimization algorithm based on optimized long short-term memory suggested by \cite{li2020hybrid}. \citet{shi2013hybrid} carried out radial basis function neural network and the Least Square Support Vector Machine (LSSVM) based hybrid wind speed forecasting model for the very short-term horizon. VMD-DE-ESN (Variational Mode Decomposition associated Differential Evolution algorithm based optimized Echo State Network) hybrid wind sped forecasting model pointed out by \cite{hu2021wind}. SDAE (Stacked Denoising Autoencoder) and SAE (Staked Autoencoder) associated with a rough extension-based Deep neural network evolved to forecast the wind speed in ultra-short-term and short-term time horizons performed in \cite{khodayar2017rough}. Using variational recurrent autoencoder incurred a generative probabilistic approach to forecast the wind speed performed in \cite{zheng2021generative}. \citet{li2020ultra} suggested Two LSTM (long short-term memory) networks are associated with MM (mathematical morphology) decomposer-based wind speed forecasting in the ultra-short-term. \citet{afrasiabi2020advanced} designed a probabilistic forecasting of wind speed using DMDNN (deep mixture density neural network). The combinational model (VMD-ConvLSTM-ES) association of VMD (Variational Mode Decomposition), ConvLSTM (Convolutional Long-Short Memory Network), and error scheme analysis-based wind power forecasting evolved in \cite{sun2020short}. The ICEEMDAN-GRU-ARIMA-ECM (Improved Complete Ensemble Empirical Mode Decomposition with Adaptive Noise-Gated recurrent unit-autoregressive integral moving average-Error Correction Method) based hybrid wind speed forecasting model for short-term horizon is carried out by \cite{duan2021short}. The limitation of SVR is the parameter tuning is not satisfactory. SVM (Support Vector Machines) utilizes the kernel to map the nonlinear problem into high-dimensional space, but is not capable enough to handle the colossal dataset; the computation takes a long time. LSTM overcame the drawbacks of the RNN, such as gradient issues (vanishing). Still, it cannot process the data in parallel; transfer learning is impossible, much memory is needed, extremely far long-term dependencies are not handled effectively, and convergence takes much time. The performance of the LSTM  can be improved by including the attention mechanism. Deep learning models can handle big data and extract significant input features, but are not good enough to address long-term dependence. Dilated CNN (Convolutional Neural Network) is good, but the transformer network is impressive and promising. Recently, transformer networks have been used for time series forecasting \cite{wu2020deep}. 
\par The inferences from the literature review are combinations of forecasting models, and hybrid models are the latest trends that significantly improve forecasting performance. Past wind speed knowledge is utterly essential for developing univariate forecasting models. Adding more irrelevant or less correlated inputs makes the model diverge from the convergence. Large training times and entrapped into local minimum are significant issues regarding the deep learning neural network based on forecasting models. Because of these limitations, the accuracy is decaying. 
\par With the motivational inspiration from the literature review, this paper is devoted to developing the enhanced wind speed forecasting model using a robust transformer network, ICEEMDAN, multi-layer perceptron neural network, and error correction strategy. One of the most significant challenges of sequence learning-based models (RNN, LSTM, and so on) are handling complex and dynamic wind data. However, the proposed model is developed based on the transformer network, which can overcome this challenge.\\
Novel contributions to the proposed astute hybrid wind speed forecasting model are: 
\begin{enumerate}
\item	The novel idea is to design the astute wind speed forecasting model using the transformer network, ICEEMDAN, and residual error correction strategy for the first time.  
\item	The procedure used for forecasting wind speed is new. Forecaster realization capability and dynamic handling capability were improved using the decomposition (ICEEMDAN) method.
\item Transformer networks are used for each decomposed subseries forecasting, and the wind speed forecast is performed by fusion.
\item 	Enhance the wind speed forecasting accuracy with a lesser performance index by residual error correction strategy.
\item The different hub height-based acquired real-time wind farm datasets were used for performance validation.
\item Carry out different horizons based on wind speed forecasting.
\end {enumerate}
\section{TECHNICAL CONCEPT AND FRAMEWORK OF THE PROPOSED HYBRID WIND SPEED FORECASTING MODEL DESIGN}
\begin {figure}
\centering
\includegraphics [width=0.5\textwidth]{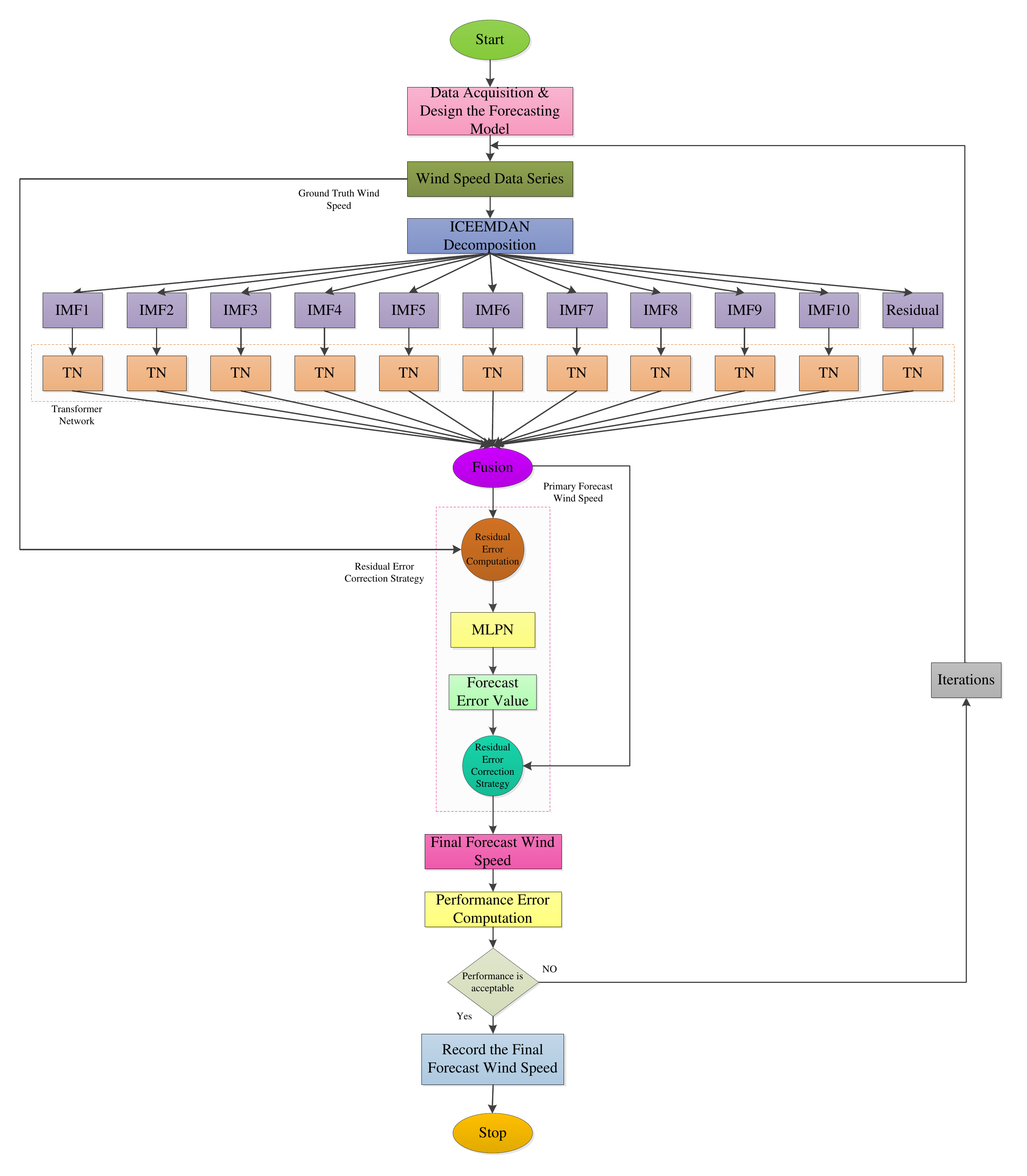}
\caption{The framework of the proposed hybrid wind speed forecasting model. The acquired wind farm data is split into two sets, one is used for the training phase, and the other is used for the testing phase. Then, decompose the wind speed series into subseries (IMFs and residuals) using ICEEMDAN. Each subseries is fed as input to the transformer network (TN) and performs the forecasting of the subseries. The fusion of all forecast subseries is done, reconstructed, and obtained the primary forecast wind speed. For each data point based residual error is computed that is fed as input to the MLPN network to perform the future residual error values forecasting because it can learn any complex problem. Finally, the forecast wind speed values are corrected by adding the forecast residual error values to the primary forecast wind speed. Thereafter, the performance is verified using performance error index computation. If acceptable, record the value, else do so until reaching the stopping criteria (max iterations).}
\centering
\label{Figure 1}
 \end{figure}

The issue of understanding deep underlying patterns, instability, overfitting, fall in local minima, selection of optimal hyperparameters are some limitations of the artificial neural network-based forecasting model. A deep neural network mitigates the issue of handling underlying deep input patterns. Other than shallow feed-forward networks, deep learning neural networks and the recurrent neural network (LSTM) receive particular importance in forecasting applications because of memory storage and recurrent characteristics. Still, exploding gradients and vanishing gradients are the two issues that worsen the model's performance. LSTM and GRU are special variants of the recurrent neural network to address the limitations of the RNN. Compared to LSTM, GRU is faster in terms of training speed because of the fewer parameters than LSTM. CNN is generally used for image processing applications, but some researchers perform CNN-LSTM based hybrid wind speed forecasting. The advantage of CNN is better feature extraction capability. Transformer networks can improve the ability to handle diversified and complex wind data with the help of self-attention mechanisms. The core idea of this paper is the accurate forecasting of wind speed using an astute hybrid model with the association of decomposition (ICEEMDAN), transformer network, fusion, multilayer perception neural network, and residual error correction methods (ICEEMDAN-TNF-MLPN-RECS). The framework of the proposed hybrid wind speed forecasting model is shown in Figure \ref{Figure 1}.  

\subsection{Data Source}
\begin {figure}
\centering
\includegraphics[width=0.8\linewidth,height=0.25\textheight]{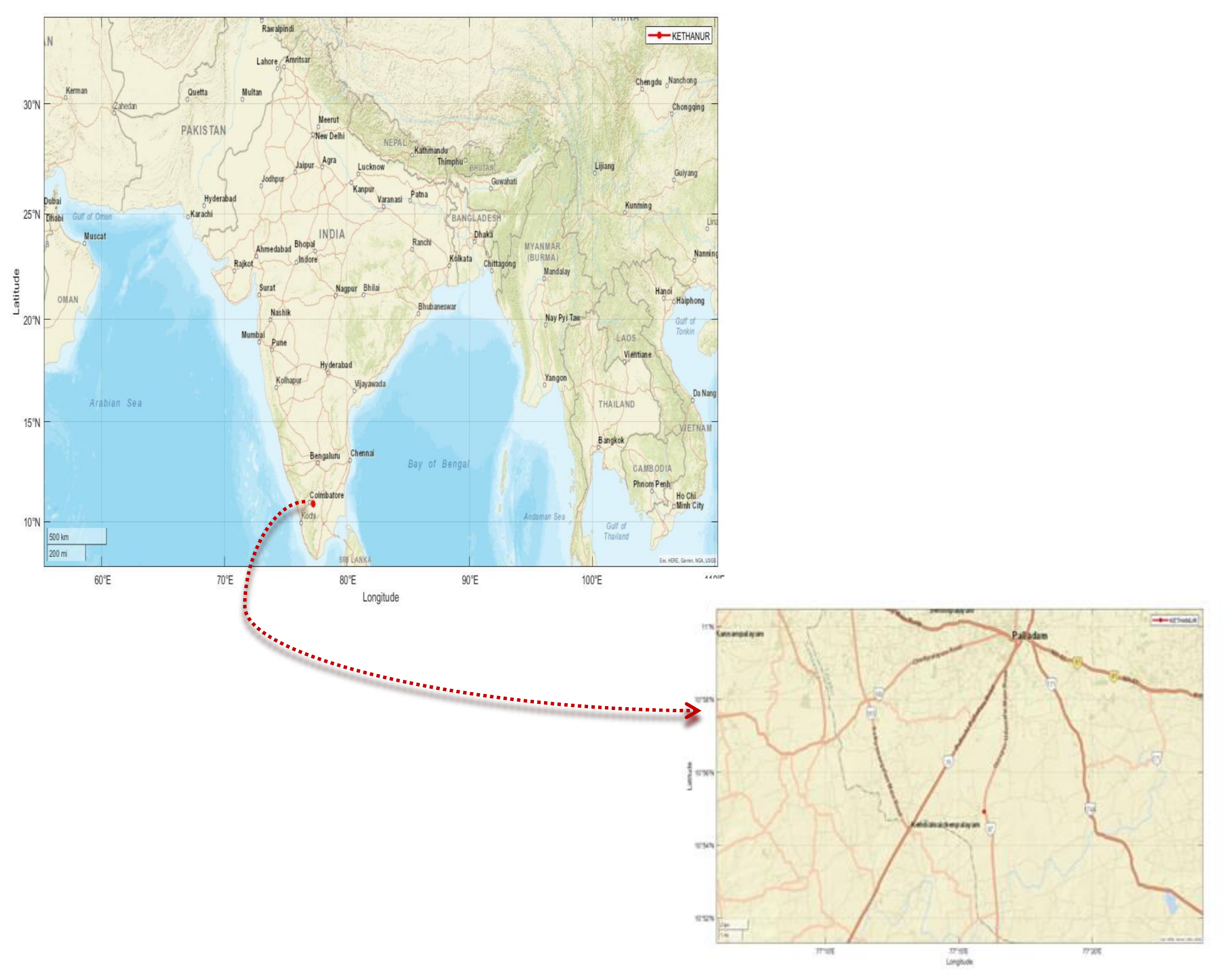}
\caption{Wind farm dataset collected location. The star in red color shows the wind farm location, Kethanur (latitude of 10.9153 N and longitude of 77.2657 N).}
\centering
\label{Figure 2}
 \end{figure}

The wind farm data used for this exploration is collected from the Kethanur location wind farm with different hub heights of 65 meters (m) and 80 meters (m). Wind farm data collected with location latitude of 10.9153 N and longitude of 77.2657 N. Geo map of wind farm data collected location shown in Figure \ref{Figure 2}. Case study 1 is Kethanur wind farm 65 m wind turbine hub height based on retrieved real-time wind farm data, and case study 2 is Kethanur wind farm 80 m wind turbine hub height based on retrieved real-time wind farm data. 
Each case study's wind farm data comprises 1,05,120 data points measured during the period of two years (2018-2019), which is 10 minutes averaged measured (ground truth) wind speed. We split wind speed data samples for training and testing for two case studies. We trained the proposed model using one-year (2018-year 10-minute average wind speed) wind farm real-time data comprising 52,560 data points. We tested the trained model performance using a one-year (2019-year 10-minute average wind speed) dataset containing 52,560 data points.

\subsection{Performance Error Index}
 Proposed model performance and comparative analysis were evaluated and verified using the following five standard performance error indices such as MAE-Mean Absolute Error, MAPE-Mean Absolute Percentage Error, MRE-Mean Relative Error, MSE-Mean Squared Error, and RMSE-Root Mean Square Error \cite{wilks2011statistical}.
\begin{equation} \label{EQ__1_} 
RMSE=\sqrt{\frac{1}{T} \left(\sum _{u=1}^{T}(G_{u} -F_{u} )^{2}  \right)}  
\end{equation} 
\begin{equation} \label{EQ__2_} 
MAPE=\left(\frac{1}{T} \sum _{u=1}^{T}\left(\frac{\left|G_{u} -F_{u} \right|}{\left|G_{u} \right|} \right)*100 \right) 
\end{equation} 
\begin{equation} \label{EQ__3_} 
MAE=\left(\frac{1}{T} \sum _{u=1}^{T}\left(\left|G_{u} -F_{u} \right|\right) \right) 
\end{equation} 
\begin{equation} \label{EQ__4_} 
MSE=\frac{1}{T} \left(\sum _{u=1}^{T}(G_{u} -F_{u} )^{2}  \right) 
\end{equation} 
\begin{equation} \label{EQ__5_} 
MRE=\left(\frac{1}{T} \sum _{u=1}^{T}\left(\frac{\left|G_{u} -F_{u} \right|}{\left|G_{u} \right|} \right) \right) 
\end{equation} 
Let, $G_{u} $- Ground truth value of data points `u', $F_{u} $- Forecasting value of data points `u', and $\left|G_{u} \right|$- Absolute ground-truth value of data points `u'.

\subsection{Proposed Astute Hybrid Approach Procedural Steps}
Procedural steps of the proposed astute hybrid wind speed forecasting model are as follows: 
\begin{steps}
\item  Start the forecasting model design process.  
\item  Perform data collection of measured wind speed and meteorological weather data from the Kethanur location at two hub heights. 
\item  Divide the collected dataset into training data and testing data. 
\item  Decompose the wind speed using ICEEMDAN, obtain the 10 IMFs and residual.  
\item  For each IMFs and residual, an individual transformer network model is used to forecast the future IMFs and residue. 
\item  The forecast IMFs and residue are composed, reconstructed, and obtained the primary forecast wind speed in the fusion stage.
\item  Compute the residual error using the equation \eqref{EQ__39_} 
\item  Perform error value forecasting using a multi-layer perceptron neural network. 
\item  The forecasting wind speed value is corrected by adding the forecast residual error to each data point's primary forecast wind speed value. 
\item  Compute the final performance error-index evaluation using the equations (\ref{EQ__1_}-\ref{EQ__5_}). Continue the process until reach the stopping condition.
\item  Record the value and End. 
\end {steps}
\subsection{Improved Complete Ensemble Empirical Mode Decomposition with Adaptive Noise}
Expeditious variation occurs in wind speed; the decomposition method helps to minimize the stochastic effect. The feature extraction capability is improved by employing the decomposition method. EMD and its other variants are having issues like reconstruction errors, residual noise present in the modes, and the occurrence of spurious modes. These limitations are subdued by ICEEMDAN \cite{colominas2014improved}. Wind speed series uncertainty and noise were reduced by utilizing the ICEEMDAN (improved complete ensemble empirical mode decomposition with adaptive noise). The subseries (IMFs modes-sum of amplitude, frequency modulation functions, and residue-monotonic trend) are derived from the original raw real-time wind speed. The relative subseries (IMFs) are obtained from the wind farm data (irregular wind speed series) from ICEEMDAN. The proposed model uses the ICEEMDAN for the decomposition, which overcomes the limitation concerned with the other decomposition methods and reconstructs the original time series without any added noise. Forecasting performance improves effectively by the mixture model using ICEEMDAN, error correction, and transformer model. 
\par Algorithm 1 (ICEEMDAN): The input wind speed series  sampling sequence of the wind speed series is ‘m’.
\begin{procedure}
\item Perform a realization of local mean computation $i^{(j)} =i+\alpha _{0} K_{1} \left(z^{(j)} \right)$and generate the initial first residue according to equation \eqref{EQ__6_}.
\begin{equation} \label{EQ__6_} 
\phi _{1} =\left\langle L(i^{(j)} )\right\rangle  
\end{equation} 
\item Compute the first mode during the initial stage (j=1)$\tilde{y}_{1} =i-\phi _{1} $.
\item Determine the second mode using equation \eqref{EQ__7_} with the help of the computed second residue, which is the average of the local mean of the realization $\phi _{1} +\alpha _{1} K_{2} \left(z^{(j)} \right)$.
\begin{equation} 
\label{EQ__7_} 
\tilde{y}_{2} =\phi _{1} -\phi _{2} =\phi _{1} -\left\langle L\left(\phi _{1} +\alpha _{1} K_{2} \left(z^{(j)} \right)\right)\right\rangle  
\end{equation} 
\item Compute the j${}^{th}$ residue for j=3,4, {\dots}, J.
\begin{equation} 
\label{EQ__8_} 
\phi _{k} =\left\langle L\left(\phi _{j-1} +\alpha _{j-1} K_{j} \left(z^{(j)} \right)\right)\right\rangle                                                  
\end{equation}
\item Determine${}^{\ }$j${}^{th}$${}^{\ }$ mode using equation \eqref{EQ__9_}.

\begin{equation} 
\label{EQ__9_} 
\tilde{y}_{j} =\phi _{j-1} -\phi _{j} 
\end{equation}
\item Increments the j value and repeats Procedure 4 until it reaches J.
\end {procedure}
\par Let, ${\alpha _{j} =\varepsilon _{j} std\left(\phi _{j} \right)}$-Constant selected to get an expected signal to noise ratio between the included noise and residue,${std\left(.\right)}$${}{-\ }$standard deviation,${}{\varepsilon _{o} }$-reciprocal of the expected SNR (signal-to-noise ratio) between the first included noise, and the considered series, $z^{(j)} $ - zero-mean white Gaussian noise and unity variance, $\left\langle .\right\rangle $- during realization mean commutating process, $L\left(.\right)$- local mean of the wind speed series, $K_{j} \left(.\right)$- using the EMD generated $j^{th} $  modes.

The parameter of the ICEEMDAN is realization number = 90, maximum shifting iteration = 1000, the ratio of the standard of added noise = 0.3.  
\subsection{Transformer Network}
The concept of self and mutual attention incorporated into the deep learning neural network makes a transformer neural network. Attention mechanism neural network architecture resembles retrieval values, like humans. Attention mechanisms can get far away tokens of pertinent information because based on the relevancy, the attention weights are assigned with the help of the previous state. At the same time, parallelly process all tokens and compute the attention weights \cite{vaswani2017attention}.
\begin {figure}
\centering
\includegraphics [width=0.5\textwidth]{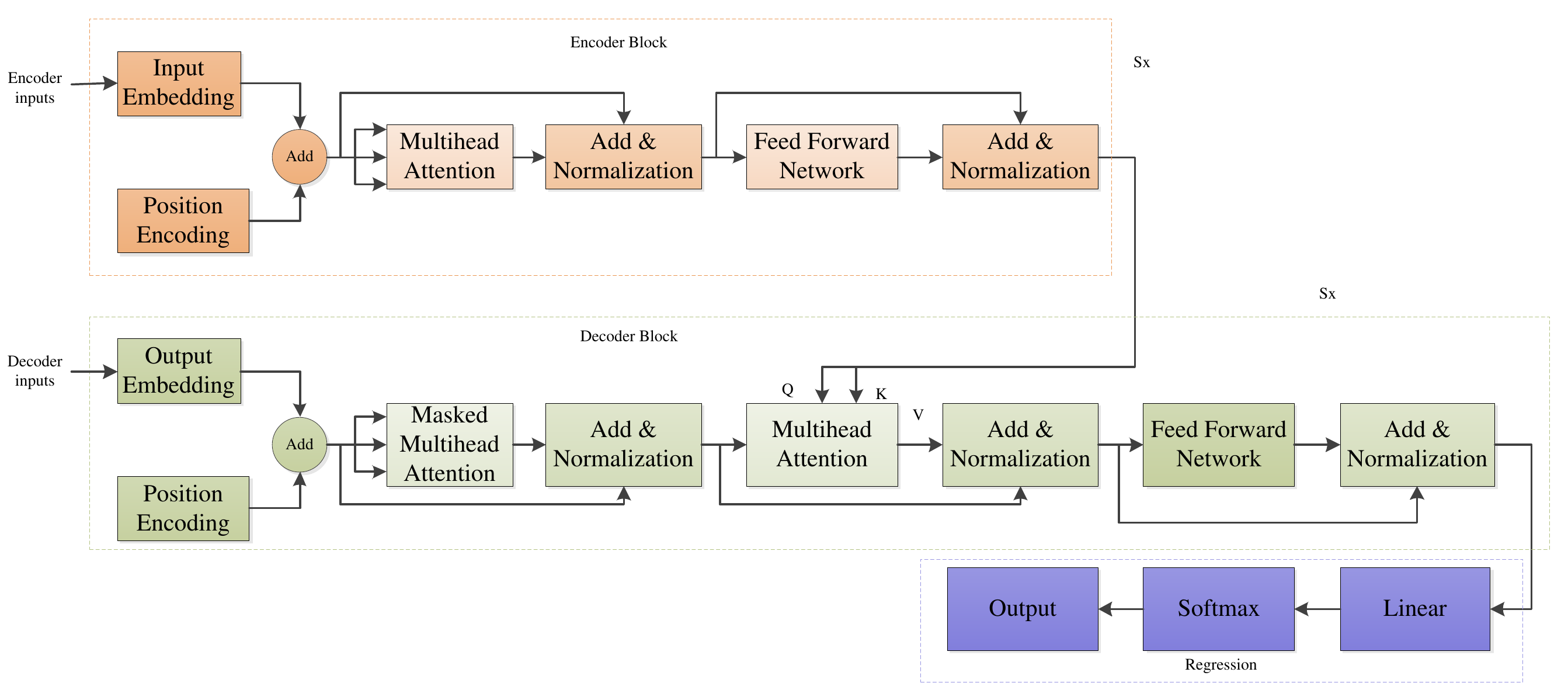}
\caption{A detailed framework of the transformer network. Encoder and decoder connected through attention. The encoder doesn't care about the sequence order, but position embedding takes care of the sequence order. Encoder layer self-attention sub-layer, fully connected feed-forward sub-layer. Normalization is between each sub-layer. The output of the encoder is a dimension vector. A decoder comprises an input layer, masked self-attention, feed-forward sub-layer, and output layer.}
\centering
\label{Figure 3}
\end{figure}
A detailed framework of the transformer network is shown in Figure \ref{Figure 3}. Advantages of transformer networks are 1. Parallel computation, 2. Draw a connection between any part of the sequence, 3. New embedding and 4. No problem concerning long-term dependencies.
\par The framework of the transformer network is formed with the encoder and decoder structure; the relevancy among the inputs is obtained in each input layer that further feeds as the inputs to the successive encoding layers. In an extended sequence, new tokens currently generated relevancy to earlier tokens are measured using a self-attention score. The high relevancy input part has a high self-attention score, and the low relevancy input part has a low self-attention score. The temporal dependency handling capability of the transformer network improved, including the positional information in inputs, thus aid to the self–attention mechanism. Multiheaded attention is the core of the transformer network; the output of the encoding block is fed as the input to the multiheaded attention of the decoder block. Decoder and encoder blocks repeated for a couple of the time of a couple of layers. The order of the data point concerning the time can be handled by the position embedding, which helps to aware the transformer network for the order positions. The entire series simultaneously passed to the network; thus, no issue of gradient vanishing and exploding. The prominent features of the inputs are weighted differentially, and they can handle the big data. Fusion techniques are used to reduce the impact of the lousy forecasting model. Self-attention mechanisms aid in filtering out the non-relevant input and give much importance to the parts of the more relevant input. Self-attention leverage of challenge of the parallel process of all hidden state time stamps, the current token encoding is highly dependent on past token relevancy. 
\par Three parameters that are especially essential for the transformer network are 1. Query, 2. Key, 3. Value. Each token has three weight matrices that are learned during the training and obtain the query, key, and value vector for each token. The feed-forward neural network performs like a 1D convolution, which speeds up the computation. Transformer normalization aids faster training, reduces the covariate shift. Adding aid in keeping the position information and strengthening gradients. It looks similar to wind's previous history of predicting acute forecasting using layer and layer of attention. Encoder connected through an attention mechanism in which part of the sequence is significant to the current forecasting is done by the attention mechanism. Transformer-based fusion was carried out, which minimizes the training time because of the parallelization. The encoding process uses the sine function and cosine function. Position offset and masking to look ahead to later data samples are guaranteed genuine forecasting. The parameters of the transformer network are Optimizer = Adam, Encoder = I${}_{1}$, I${}_{2}$, {\dots}.., I${}_{52}$${}_{,}$${}_{560}$, Decoder input = I${}_{52}$${}_{,}$${}_{561}$, I${}_{52}$${}_{,}$${}_{562}$, {\dots}.., I${}_{10}$${}_{,}$${}_{5120}$${}_{,\ }$ Decoder output = O${}_{1}$,O${}_{2}$, {\dots}.., O${}_{52}$${}_{,}$${}_{560}$, Drop out  = 0.2, number of iteration = 10, Learning rate = 0.001, Number of stack (s) = 3, Mini batch size = 256, Number of heads = 4, Max gradient norm = 0.01.

Mathematical modeling of transformer networks:
\begin{equation} 
\label{EQ__10_} 
Input embedding =i_{T} =EE_{T} +PE_{T}                              
\end{equation}
Add and Normalize:
Add, 
\begin{equation} 
\label{EQ__11_} 
\psi =I+O
\end{equation}
\begin{equation} 
\label{EQ__12_} 
Normalize, \eta =\frac{1}{lT} \sum _{t=1}^{l}\sum _{u=1}^{T}\psi _{tu}                          
\end{equation}
\begin{equation} \label{EQ__13_} 
\sigma ^{2} =\frac{1}{lT} \sum _{t=1}^{l}\sum _{u=1}^{T}(\psi _{tu} -\eta )^{2}    
\end{equation} 
\begin{equation} \label{EQ__14_} 
Z_{ut} =\frac{\psi _{tu} -\eta }{\sqrt{\sigma ^{2} +\varepsilon } }  
\end{equation} 
Self-attention:
\begin{equation} \label{EQ__15_} 
I=\left[i_{1}^{} \quad i_{2} \quad ...\quad i_{T} \right] 
\end{equation} 
\begin{equation} \label{EQ__16_} 
O=\left[o_{1}^{} \quad o_{2} \quad ...\quad o_{T} \right] 
\end{equation} 
\begin{equation} \label{EQ__17_} 
Q=\left[q_{1}^{} \quad q_{2} \quad ...\quad q_{T} \right] 
\end{equation} 
\begin{equation} \label{EQ__18_} 
K=\left[k_{1}^{} \quad k_{2} \quad ...\quad k_{T} \right] 
\end{equation} 
\begin{equation} \label{EQ__19_} 
V=\left[v_{1}^{} \quad v_{2} \quad ...\quad v_{T} \right] 
\end{equation} 
\[Q=SW_{Q} I, K=SW_{K} I, V=SW_{V} I\] 
\begin{equation} \label{EQ__20_} 
Z=\frac{K^{Transpose} Q}{\sqrt{l} }  
\end{equation} 
Weighted average based new embedding 
\begin{equation} \label{EQ__21_} 
O=VSW 
\end{equation} 
Multihead attention:
Concatenated output vector
\begin{equation} \label{EQ__22_} 
\tilde{O}=\left[\begin{array}{l} {O_{1} } \\ {O_{2} } \\ {} \\ {O_{H} } \end{array}\right] 
\end{equation} 
\begin{equation} \label{EQ__23_} 
O=SW^{0} \tilde{O} 
\end{equation} 
Masked Self Attention:
\begin{equation} \label{EQ__24_} 
Q=SW_{Q} I 
\end{equation} 
\begin{equation} \label{EQ__25_} 
K=SW_{K} I 
\end{equation} 
\begin{equation} \label{EQ__26_} 
V=SW_{V} I 
\end{equation} 
\begin{equation} \label{EQ__27_} 
Z=\frac{K^{Transpose} Q}{\sqrt{l} }  
\end{equation} 
\setlength{\tabcolsep}{2pt}
\begin{multline}\label{EQ__28_} 
SW=\\
\begin{bmatrix}
\begin{tabular}{ccccc}
1 & $sm_{1}(Z_{12} ,Z_{22})$ & $sm_{1}(Z_{13} ,Z_{23} ,Z_{33} )$ & ... & $sm_{1}(Z_{14}, Z_{24},...Z_{ut})$\\
0 & $sm_{2}(Z_{11} ,Z_{22})$ & $sm_{2}(Z_{13} ,Z_{23} ,Z_{33})$ & ... & $sm_{2}(Z_{14}, Z_{24},...Z_{ut})$ \\
0 & 0 & $sm_{3}(Z_{13} ,Z_{23} ,Z_{33})$ & ... & $sm_{3}(Z_{14}, Z_{24},...Z_{ut})$ \\
. & . & . & . & . \\
0 & 0 & 0 &  ... & $sm_{t}(Z_{14}, Z_{24},...Z_{ut})$
\end{tabular}
\end{bmatrix}
\end{multline} 
\begin{equation} 
\label{EQ__29_} 
Weighted average based new embedding O=VSW
\end{equation}
Encoder decoder self-attention:
\begin{equation} \label{EQ__30_} 
I_{D} =\left[i_{1}^{D} \quad i_{2}^{D} \quad ...\quad i_{TD}^{D} \right] 
\end{equation} 
\begin{equation} \label{EQ__31_} 
I_{E} =\left[i_{1}^{E} \quad i_{2}^{E} \quad ...\quad i_{TE}^{E} \right] 
\end{equation} 
\begin{equation} \label{EQ__32_} 
Q=SW_{Q} I_{D}  
\end{equation} 
\begin{equation} \label{EQ__33_} 
K=SW_{K} I_{E}  
\end{equation} 
\begin{equation} \label{EQ__34_} 
V=SW_{V} I_{E}  
\end{equation} 
\begin{equation} 
\label{EQ__35_} 
Weights, Z=\frac{K^{Transpose} Q}{\sqrt{l} }                           
\end{equation}
\begin{equation} \label{EQ__36_} 
SW=soft\max (Z) 
\end{equation} 
\begin{equation} 
\label{EQ__37_} 
Weighted average based new embedding O=VSW
\end{equation}
\par Let, $sm$-softmax, $SW$- Synaptic weight, $I$- encoder input, $O$- decoder outputs, $I_{E} $- encoder output, $I_{D} $- Decoder inputs, $T$- number of inputs, $H$- number of head, $l$- length of embedding, $K^{Transpose} $- transpose of key. $EE_{u} $- encoder embedding, $PE_{u} $- position embedding, $\eta $= average, $\sigma ^{2} $- sample variance, $Z_{ut} $- normalized output, $SW^{0} $- dimensionality reduction matrix, $\varepsilon $ -numerical stability constant, Q- query, V- value, and K- key.
\subsection{Multilayer Perceptron Neural Networks}
Multilayer perceptron neural networks (MLPN) can solve complex problems with the help of nonlinear transfer function and hidden layers. MLPN consists of an input, hidden, and output layer. The number of hidden neurons is selected based on the criteria [11]. Multilayer perceptron parameters are learning rate = 0.01, Maximum iteration = 1000, learning algorithm Marquardt Levenberg learning algorithm, hidden layer = 1, activation function (f) = hyperbolic tangent sigmoid activation function, hidden neuron = 2, threshold =1, Input = residual error value, and output = forecast residual error value.
\begin{equation} \label{EQ__38_} 
MLPN_{output} =f\left(\sum _{h=1}^{2}Y_{h} SW_{h}  \right) 
\end{equation} 

Let, $Y_{h} =f\left(\right. \sum _{h=1}^{2}MLPN_{input}  SW_{input,h} \left. \right)$, $MLPN_{input} =input\; vector$, $MLPN_{output} =outputVector$, $SW_{input,h} =synaptic\; weight\; between\; input\; to\; hidden\; layer$, $SW_{h} =synaptic\; weight\; between\; hidden\; to\; output\; layer$.
\subsection{Residual Error Correction Strategy}
Eventually, the forecasting ability of the proposed hybrid model is further improved by incorporating the residual error correction strategy. \begin{equation} \label{EQ__39_} 
Residual\; Error_{u} =\left(G_{u} -F_{u}^{P} \right) 
\end{equation} 

Residual error correction, 
\begin{equation} \label{EQ__40_} 
Final forecast wind speed = F_{u}^{P} +RE_{u}^{F} 
\end{equation} 
\par $G_{u} $- Ground truth value of data points `u', $F_{u}^{P} $- Primary Forecasting value of data points `u', $Residual\; Error_{u} $ - Residual error of data points `u', $RE_{u}^{F} $-MLPN based forecast residual error value of data point `u'.
\section{PROPOSED FORECASTING MODEL EXPERIMENTAL RESULT ANALYSIS AND DISCUSSION }
The experimentation of the proposed model and other literature models are carried out in the system configuration x64-based PC, Processor	Intel(R) Xeon(R) Gold 5120 CPU @ 2.20GHz, 2195 Mhz, 14 Core(s), 28 Logical Processor(s), RAM 96.00 GB, GPU 16 GB NVIDIA QUADRO RTX 5000. We carried out the proposed hybrid model simulation on two wind farm datasets from Kethanur, India, collected at different hub height of 65 and 80 meters. The proposed hybrid forecasting model was implemented using the PyTorch framework in Python programming and environment. 
Wind speed forecasting has become pivotal for integrating the wind energy system into the modern system. Wind owns the nature of highly varying and intermittent conditions, making it tough to forecast wind speeds accurately. Nonlinear correlation exists in the wind speed identified using the intrinsic mode function, each IMF based forecasting is carried out using the individual transformer network, and the primary forecasting of wind speed is achieved by composing all outputs of the individual transformer network model. Multilayer perceptron network is employed to forecast the residual error values. The final forecast wind speed was computed using the residual error correction strategy. Forecasting accuracy is enriched by employing the residual error correction strategy. The proposed hybrid approach-based framework overcomes the limitation of GRU (Gated Recurrent Unit), LSTM, processes the input data in parallel, and improves the training time. The proposed ICEEMDAN-TNF-MLPN-RECS experimental results for case studies 1 and 2 are as follows:
\begin{table}[h]
\centering
\caption{PROPOSED ICEEMDAN-TNF-MLPN-RECS MODEL EXPERIMENTAL RESULT. The proposed wind speed forecasting model effectiveness is verified in two case studies. Both case studies suggested that model-based forecasting achieves the minimal performance error index.}
\label{Table 1}
\begin{tabular}{|l|ll|}
\hline
\multirow{2}{*}{\textbf{Performance Error Index}} & \multicolumn{2}{l|}{\textbf{Proposed   ICEEMDAN-TNF-MLPN-RECS}} \\ \cline{2-3} 
                                         & \multicolumn{1}{l|}{\textbf{Case Study 1} }    & \textbf{Case Study 2}    \\ \hline
MAE                                      & \multicolumn{1}{l|}{1.7096× 10$^{-07}$}    & 6.1565× 10$^{-07}$   \\ \hline
MAPE                                     & \multicolumn{1}{l|}{2.8416× 10$^{-06}$}    & 9.5005× 10$^{-06}$   \\ \hline
MRE                                      & \multicolumn{1}{l|}{2.8416× 10$^{-08}$}    & 9.5005× 10$^{-08}$   \\ \hline
MSE                                      & \multicolumn{1}{l|}{5.0206× 10$^{-14}$}    & 8.9289× 10$^{-13}$   \\ \hline
RMSE                                     & \multicolumn{1}{l|}{2.2407× 10$^{-07}$}    & 9.4493× 10$^{-07}$   \\ \hline
\end{tabular}
\end{table}

\begin{table}[h]
\tiny
\centering
\caption{PROPOSED HYBRID MODEL PERFORMANCE ANALYSIS WITH VARIOUS IMFS. Bold represents the best results.}
\label{Table 2}
\begin{tabular}{|p{1 cm}|l|lllll|}
\hline
\multirow{2}{*}{\textbf{\makecell{ICEEMDAN \\Various   IMFs}} }& \multirow{2}{*}{\textbf{Case study}} & \multicolumn{5}{l|}{\textbf{Performance   Error Index}}                                                                                                                    \\ \cline{3-7} 
                                         &                             & \multicolumn{1}{l|}{\textbf {MAE}}           & \multicolumn{1}{l|}{\textbf{ MAPE}}          & \multicolumn{1}{l|}{\textbf {MRE}}           & \multicolumn{1}{l|}{\textbf{ MSE}}           & \textbf{RMSE}          \\ \hline
\multirow{2}{*}{5}                       & Case Study 1                & \multicolumn{1}{l|}{3.7458× 10$^{-05}$} & \multicolumn{1}{l|}{6.2262× 10$^{-04}$} & \multicolumn{1}{l|}{6.2262× 10$^{-06}$} & \multicolumn{1}{l|}{4.2154× 10$^{-09}$} & 6.4926× 10$^{-05}$ \\ \cline{2-7} 
                                         & Case Study 2                & \multicolumn{1}{l|}{4.4014× 10$^{-05}$} & \multicolumn{1}{l|}{6.7921× 10$^{-04}$} & \multicolumn{1}{l|}{6.7921× 10$^{-06}$} & \multicolumn{1}{l|}{8.3769× 10$^{-07}$} & 9.1525× 10$^{-04}$ \\ \hline
\multirow{2}{*}{6}                       & Case Study 1                & \multicolumn{1}{l|}{2.6485× 10$^{-05}$} & \multicolumn{1}{l|}{4.4022× 10$^{-04}$} & \multicolumn{1}{l|}{4.4022× 10$^{-06}$} & \multicolumn{1}{l|}{1.3276× 10$^{-09}$} & 3.6436× 10$^{-05}$ \\ \cline{2-7} 
                                         & Case Study 2                & \multicolumn{1}{l|}{3.1379× 10$^{-05}$} & \multicolumn{1}{l|}{4.8423× 10$^{-04}$} & \multicolumn{1}{l|}{4.8423× 10$^{-06}$} & \multicolumn{1}{l|}{3.1017× 10$^{-09}$} & 5.5693× 10$^{-05}$ \\ \hline
\multirow{2}{*}{7}                       & Case Study 1                & \multicolumn{1}{l|}{9.9273× 10$^{-06}$} & \multicolumn{1}{l|}{1.6501× 10$^{-04}$} & \multicolumn{1}{l|}{1.6501× 10$^{-06}$} & \multicolumn{1}{l|}{2.0044× 10$^{-10}$} & 1.4158× 10$^{-05}$ \\ \cline{2-7} 
                                         & Case Study 2                & \multicolumn{1}{l|}{1.3334× 10$^{-05}$} & \multicolumn{1}{l|}{2.0577× 10$^{-04}$} & \multicolumn{1}{l|}{2.0577× 10$^{-06}$} & \multicolumn{1}{l|}{3.0443× 10$^{-10}$} & 1.7449× 10$^{-05}$ \\ \hline
\multirow{2}{*}{8}                       & Case Study 1                & \multicolumn{1}{l|}{6.1390× 10$^{-06}$} & \multicolumn{1}{l|}{1.0204× 10$^{-04}$} & \multicolumn{1}{l|}{1.0204× 10$^{-06}$} & \multicolumn{1}{l|}{2.7710× 10$^{-10}$} & 1.6646× 10$^{-05}$ \\ \cline{2-7} 
                                         & Case Study 2                & \multicolumn{1}{l|}{7.0148× 10$^{-06}$} & \multicolumn{1}{l|}{1.0825× 10$^{-04}$} & \multicolumn{1}{l|}{1.0825× 10$^{-06}$} & \multicolumn{1}{l|}{1.5620× 10$^{-10}$} & 1.2498× 10$^{-05}$ \\ \hline
\multirow{2}{*}{9}                       & Case Study 1                & \multicolumn{1}{l|}{1.3503× 10$^{-06}$} & \multicolumn{1}{l|}{2.2444× 10$^{-05}$} & \multicolumn{1}{l|}{2.2444× 10$^{-07}$} & \multicolumn{1}{l|}{3.5546× 10$^{-12}$} & 1.8854× 10$^{-06}$ \\ \cline{2-7} 
                                         & Case Study 2                & \multicolumn{1}{l|}{2.0643× 10$^{-06}$} & \multicolumn{1}{l|}{3.1856× 10$^{-05}$} & \multicolumn{1}{l|}{3.1856× 10$^{-07}$} & \multicolumn{1}{l|}{9.9411× 10$^{-12}$} & 3.1530× 10$^{-06}$ \\ \hline
\multirow{2}{*}{\textbf{10}}                      & \textbf{ Case Study 1 }               & \multicolumn{1}{l|}{\textbf{ 1.7096× 10$^{-07}$}} & \multicolumn{1}{l|}{\textbf { 2.8416× 10$^{-06}$}} & \multicolumn{1}{l|}{\textbf { 2.8416× 10$^{-08}$}} & \multicolumn{1}{l|}{\textbf{ 5.0206× 10$^{-14}$}} & \textbf{ 2.2407× 10$^{-07}$} \\ \cline{2-7} 
                                         & \textbf{ Case Study 2}                & \multicolumn{1}{l|}{\textbf { 6.1565× 10$^{-07}$} }& \multicolumn{1}{l|}{\textbf{ 9.5005× 10$^{-06}$}} & \multicolumn{1}{l|}{\textbf{ 9.5005× 10$^{-08}$}} & \multicolumn{1}{l|}{\textbf{ 8.9289× 10$^{-13}$}} & \textbf{ 9.4493× 10$^{-07}$} \\ \hline
\multirow{2}{*}{11}                      & Case Study 1                & \multicolumn{1}{l|}{7.8790× 10$^{-06}$} & \multicolumn{1}{l|}{1.3096× 10$^{-04}$} & \multicolumn{1}{l|}{1.3096× 10$^{-06}$} & \multicolumn{1}{l|}{1.2314× 10$^{-10}$} & 1.1097× 10$^{-05}$ \\ \cline{2-7} 
                                         & Case Study 2                & \multicolumn{1}{l|}{8.4388× 10$^{-06}$} & \multicolumn{1}{l|}{1.3022× 10$^{-04}$} & \multicolumn{1}{l|}{1.3022× 10$^{-06}$} & \multicolumn{1}{l|}{2.1442× 10$^{-10}$} & 1.4643× 10$^{-05}$ \\ \hline
\multirow{2}{*}{12}                      & Case Study 1                & \multicolumn{1}{l|}{7.7075× 10$^{-06}$} & \multicolumn{1}{l|}{1.2811× 10$^{-04}$} & \multicolumn{1}{l|}{1.2811× 10$^{-06}$} & \multicolumn{1}{l|}{1.2638× 10$^{-10}$} & 1.1242× 10$^{-05}$ \\ \cline{2-7} 
                                         & Case Study 2                & \multicolumn{1}{l|}{8.4056× 10$^{-06}$} & \multicolumn{1}{l|}{1.2971× 10$^{-04}$} & \multicolumn{1}{l|}{1.2971× 10$^{-06}$} & \multicolumn{1}{l|}{2.0277× 10$^{-10}$} & 1.4240× 10$^{-05}$ \\ \hline
\end{tabular}
\end{table}
\subsection{Case Study 1: Proposed ICEEMDAN-TNF-MLPN-RECS Model Experimental Result}
\begin {figure}
\centering
\includegraphics [width=0.5\textwidth]{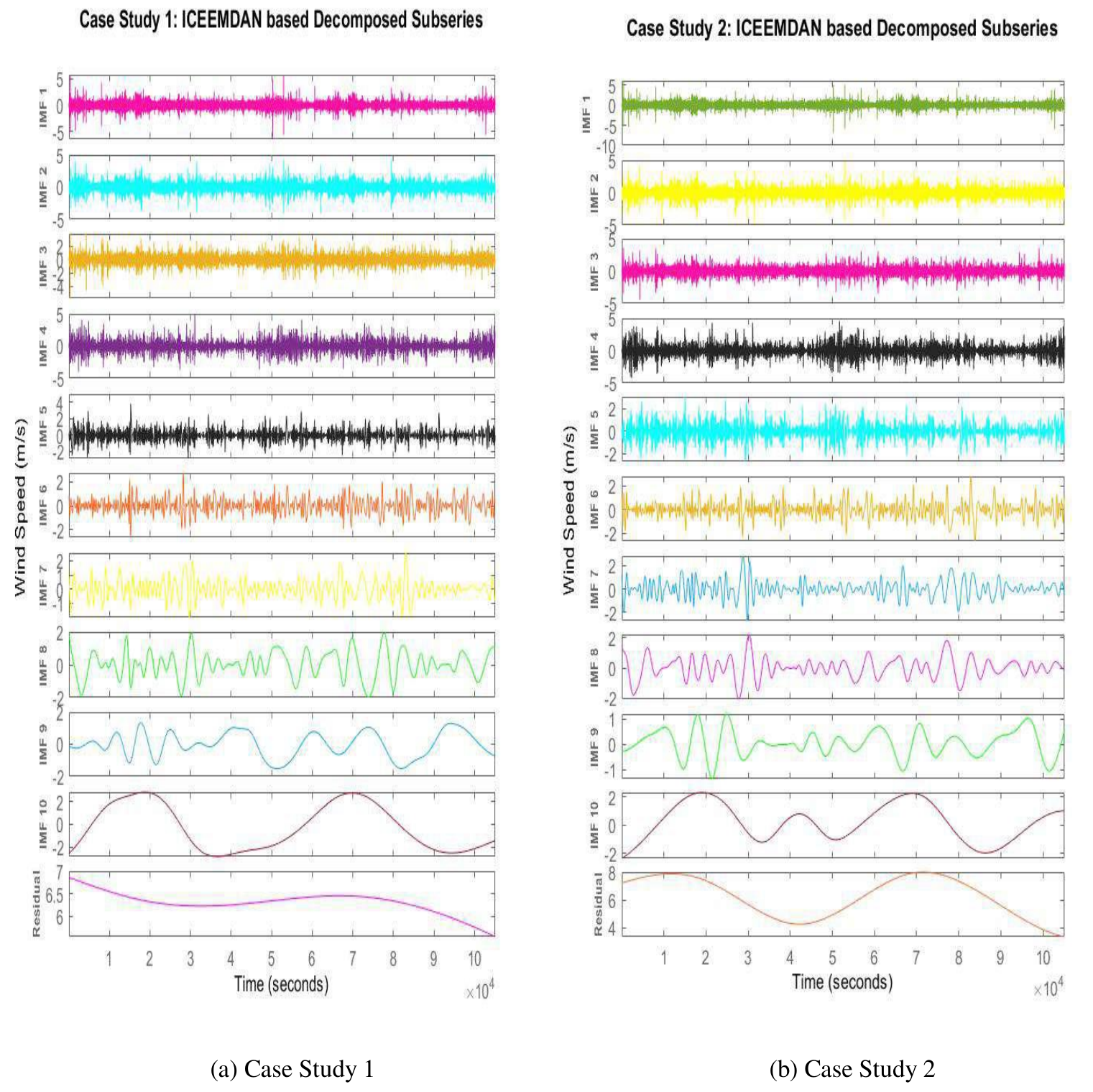}
\caption{Case Study 1 and 2: ICEEMDAN based on decomposed subseries.  ICEEMDAN decomposed the collected case study 1 and 2 whole wind speed series into ten intrinsic mode functions (IMFs) and one residual subseries.}
\centering
\label{Figure 4}
\end{figure}

\begin {figure}
\centering
\includegraphics [width=0.5\textwidth]{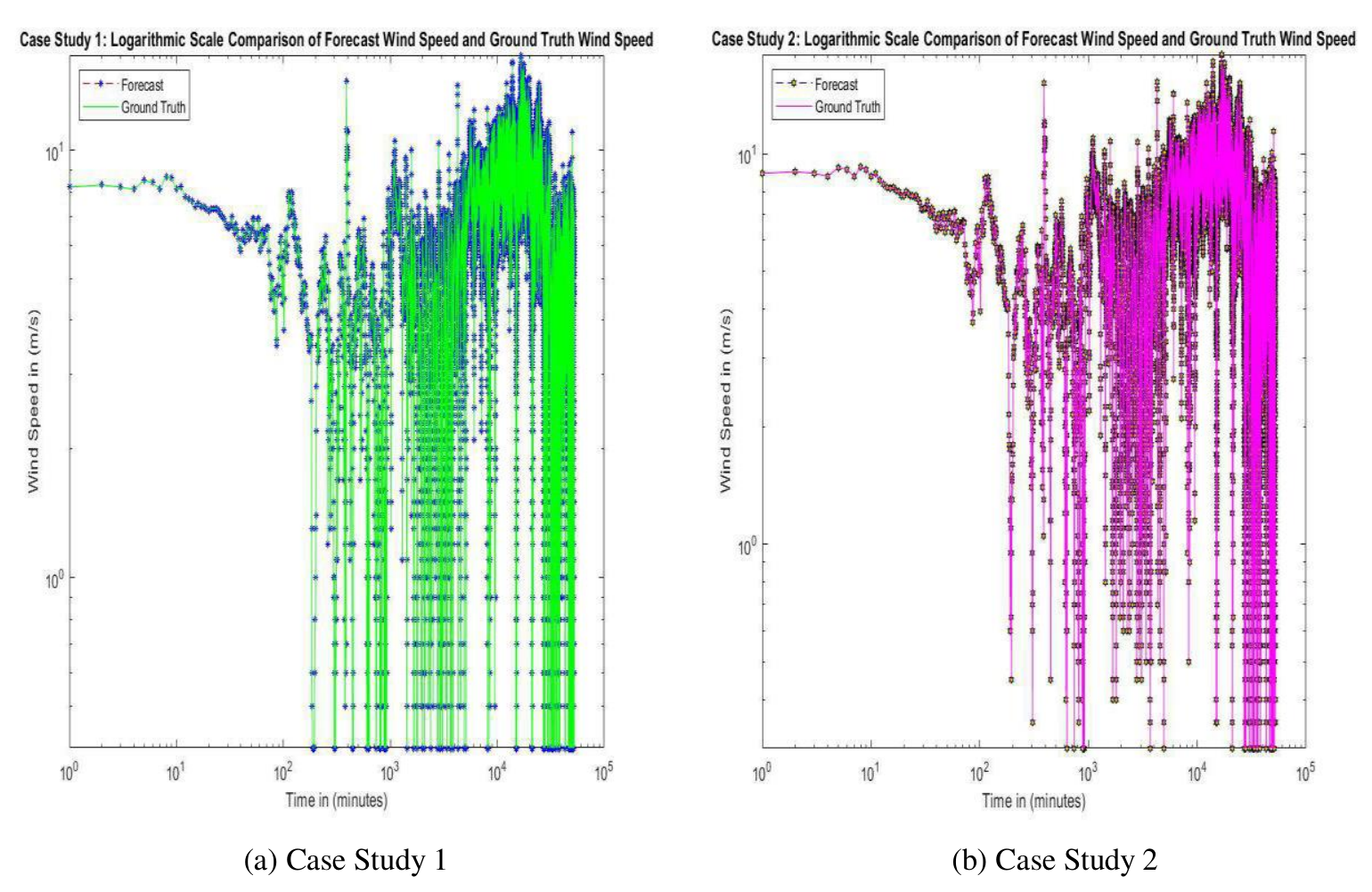}
\caption{Case Study 1 and 2: Logarithmic scale comparison of forecast wind speed and ground truth wind speed. The proposed hybrid approach-based forecast wind speed results are precise and similar to the ground truth wind speed values.}
\centering
\label{Figure 5}
\end{figure}

The real-time dataset, acquired at 65-meter hub heights from a wind farm in Kethanur, India, is used for case study 1 analysis. Decomposed case study 1 subseries were split into training and testing data sets. We use 2018 year of data samples for training purposes and 2019 year of data samples for testing purposes. The proposed hybrid wind speed forecasting model based on experimental results are tabulated in Table \ref{Table 1}, shown in Figures \ref{Figure 4} (a), and \ref{Figure 5} (a).

The proposed combined hybrid model fosters accurate wind speed forecasting with the lowest performance error-index to evaluation with case study 1 data, such as MAE=1.7096× 10$^{-07}$, MAPE=2.8416× 10$^{-06}$, MRE=2.8416× 10$^{-08}$, MSE=5.0206× 10$^{-14}$, and RMSE=2.2407×10$^{-07}$, which confirms improved forecasting accuracy of the proposed model.
\subsection{Case Study 2: Proposed ICEEMDAN-TNF-MLPN-RECS Model Experimental Result}
The real-time dataset collected at 80-meter hub heights from the wind farm in Kethanur, India, is used for case study 2 analysis. Decomposed case study 2 subseries were split into training and testing data set. We use 2018 year of data samples for training purposes and 2019 year of data samples for testing purposes. The proposed hybrid wind speed forecasting model based on the obtained experimental results is tabulated in Table \ref{Table 1}, displayed in Figures \ref{Figure 4} (b), and \ref{Figure 5} (b). Evaluation on case study 2 data, the proposed ICEEMDAN-TNF-MLPN-RECS hybrid model results in the smallest performance error index such as MAE=6.1565× 10$^{-07}$, MAPE=9.5005×10$^{-06}$, MRE=9.5005×10$^{-08}$, MSE=8.9289×10$^{-13}$, and RMSE=9.4493×10$^{-07}$, which confirms the proposed model accurate wind speed forecasting ability.
Significantly, the error values are the smallest in both case studies regarding the long-term timescale, i.e., one year ahead of wind speed forecasting, which are noticed from Table \ref{Table 1} obtained results. 



\par Furthermore, we analyzed the proposed hybrid model performance with various intrinsic mode functions (IMFs) numbers between five and twelve, and the obtained results were tabulated in Table \ref{Table 2}. Based on the analysis, we understand the significance of the IMF's number and its impact on the performance index. IMF 1, IMF 2, IMF 3, IMF 4, and IMF 5 carry essential information about the wind speed series. Therefore, the performance analysis was carried out between five and twelve. If the IMF's are more than ten have very little pertinent information, it may be redundant. Hence, it does not improve the forecasting performance in terms of the minimal performance index, as shown in Table \ref{Table 2}. Based on the Table \ref{Table 2} investigation, the IMF's number ten results in the smallest performance index than the other IMF's numbers in both case studies. In addition, we analyzed the proposed hybrid approach forecasting performance effectiveness with other state-of-the-art models and baseline models.


\section{COMPARATIVE PERFORMANCE ANALYSIS}
The strength of the transformer network is parallel computing, faster training, and being able to handle long-term dependencies. The conventional EMD method has limitations related to mode overlap and marginal impact. The above-said barrier can be overcome by ICEEMDAN based decomposition. ICEEMDAN helps to learn the wind speed series pattern with IMFs and residuals. Compensating error values are forecast using the multilayer perceptron neural network, the residual error correction strategy aids in enhancing the exactness of the forecast wind speed. By combining these models to devise an astute hybrid wind speed forecasting model. Comparative performance evaluation was carried out on two different wind turbine hub heights like 65 m and 80 m, from a real-time wind farm located in Kethanur, India. We used the parameters of the considered models for comparative analysis, as per the description mentioned in their respective papers. The performance index-based comparative summarizes for case study 1 and case study 2 are depicted in Tables \ref{Table 3} and \ref{Table 4}, respectively. The proposed forecasting model is promising and offers improved wind speed forecasting with better forecasting accuracy and negligible performance index.

\subsection{Performance Analysis for Case Study 1}
\begin {figure}
\centering
\includegraphics [width=0.5\textwidth]{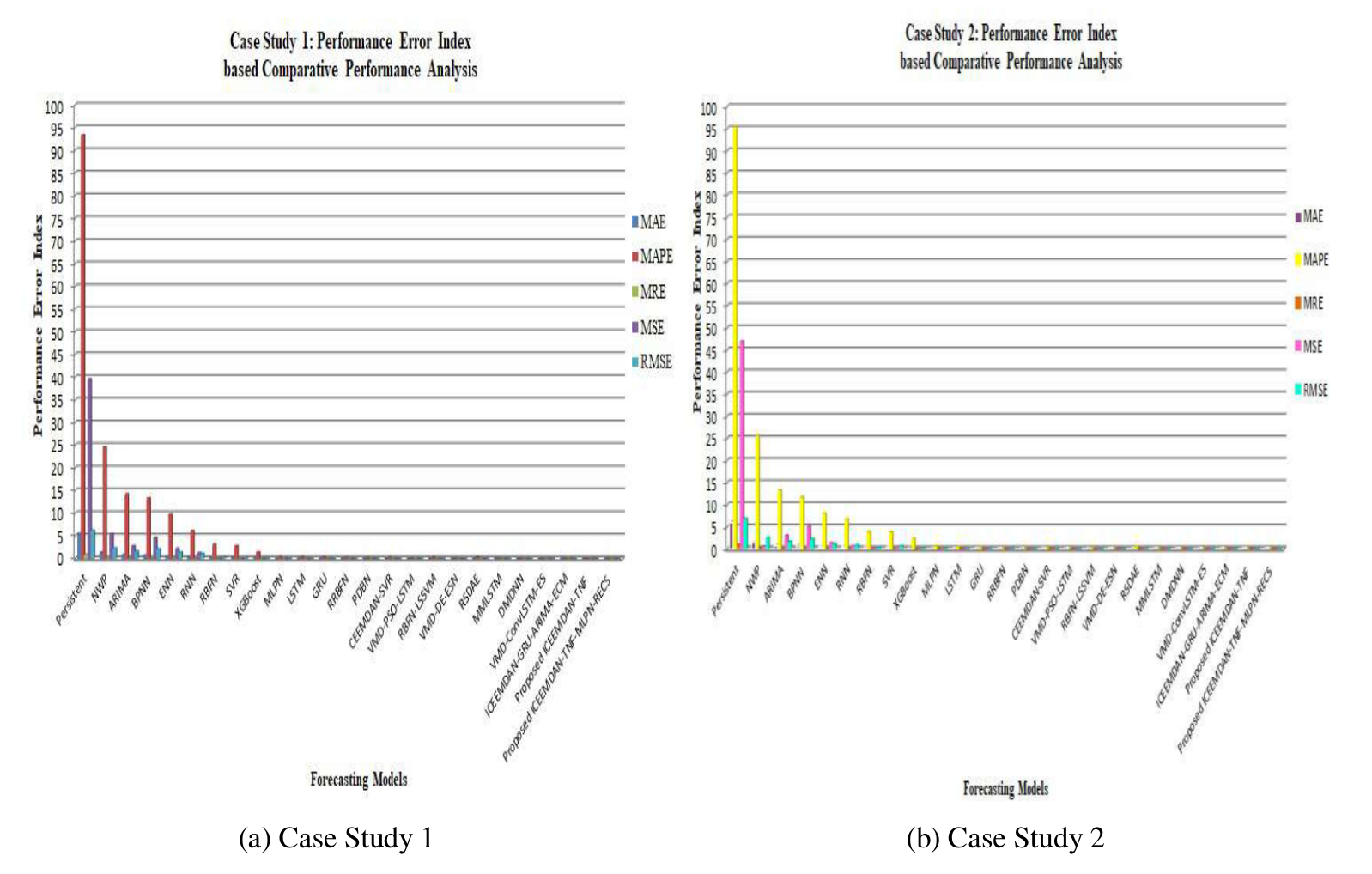}
\caption{Case Study 1 and 2: Performance error index based on comparative performance analysis.}
\centering
\label{Figure 6}
\end{figure}

From the careful investigation of Table \ref{Table 3}, we noticed that the proposed model-based simulation results proved the validity with a negligible small performance error index, compared to other models for case study 1 based performance analysis. We highlight the best models with minimum error values in bold, among other models. Figure \ref{Figure 6} (a) presents a better understanding of the performance error index-based comparative analysis.
\begin{table}[]
\tiny
\centering
\caption{COMPARATIVE PERFORMANCE RESULTS ANALYSIS FOR CASE STUDY 1. Bold represents the best results.}
\label{Table 3}
\begin{tabular}{|p{0.5cm}|p{1cm}|lllll|}
\hline
\multirow{2}{*}{\textbf{S.   NO}} & \multirow{2}{*}{\textbf{\makecell{Forecasting\\ Model}} }& \multicolumn{5}{l|}{\textbf{Performance Error Index} }                                                                                                                                  \\ \cline{3-7} 
                         &                                    & \multicolumn{1}{l|}{MAE}             & \multicolumn{1}{l|}{MAPE}            & \multicolumn{1}{l|}{MRE}             & \multicolumn{1}{l|}{MSE}             & RMSE            \\ \hline
1                        & Persistent \cite{franke1995investigations}             & \multicolumn{1}{l|}{5.6162}          & \multicolumn{1}{l|}{93.3513}         & \multicolumn{1}{l|}{0.9335}          & \multicolumn{1}{l|}{39.6054}         & 6.2933          \\ \hline
2                        & NWP  \cite{landberg1994short}                       & \multicolumn{1}{l|}{1.4847}          & \multicolumn{1}{l|}{24.6784}         & \multicolumn{1}{l|}{0.2468}          & \multicolumn{1}{l|}{5.4908}          & 2.3432          \\ \hline
3                        & ARIMA \cite{palomares2009arima}                      & \multicolumn{1}{l|}{0.8581}          & \multicolumn{1}{l|}{14.2636}         & \multicolumn{1}{l|}{0.1426}          & \multicolumn{1}{l|}{2.8602}          & 1.6912          \\ \hline
4                        & BPNN \cite{madhiarasan2016performance}                       & \multicolumn{1}{l|}{0.8022}          & \multicolumn{1}{l|}{13.3332}         & \multicolumn{1}{l|}{0.1333}          & \multicolumn{1}{l|}{4.6074}          & 2.1465          \\ \hline
5                        & ENN \cite{madhiarasan2016performance}                        & \multicolumn{1}{l|}{0.5854}          & \multicolumn{1}{l|}{9.7309}          & \multicolumn{1}{l|}{0.0973}          & \multicolumn{1}{l|}{2.2634}          & 1.5045          \\ \hline
6                        & RNN \cite{barbounis2006long}                       & \multicolumn{1}{l|}{0.3745}          & \multicolumn{1}{l|}{6.2242}          & \multicolumn{1}{l|}{0.0622}          & \multicolumn{1}{l|}{1.3127}          & 1.1457          \\ \hline
7                        & RBFN \cite{junli2010wind}                      & \multicolumn{1}{l|}{0.1897}          & \multicolumn{1}{l|}{3.1535}          & \multicolumn{1}{l|}{0.0315}          & \multicolumn{1}{l|}{0.1410}          & 0.3755          \\ \hline
8                        & SVR \cite{botha2017forecasting}                      & \multicolumn{1}{l|}{0.1718}          & \multicolumn{1}{l|}{2.8549}          & \multicolumn{1}{l|}{0.0285}          & \multicolumn{1}{l|}{0.0659}          & 0.2567          \\ \hline
9                        & XGBoost \cite{ahmadi2020long}                   & \multicolumn{1}{l|}{0.0910}          & \multicolumn{1}{l|}{1.5127}          & \multicolumn{1}{l|}{0.0151}          & \multicolumn{1}{l|}{0.0566}          & 0.2380          \\ \hline
10                       & MLPN \cite{madhiarasan2017comparative}                      & \multicolumn{1}{l|}{0.0339}          & \multicolumn{1}{l|}{0.5633}          & \multicolumn{1}{l|}{0.0056}          & \multicolumn{1}{l|}{0.0091}          & 0.0956          \\ \hline
11                       & LSTM \cite{araya2018lstm}                    & \multicolumn{1}{l|}{0.0264}          & \multicolumn{1}{l|}{0.4395}          & \multicolumn{1}{l|}{0.0010}          & \multicolumn{1}{l|}{0.0010}          & 0.0318          \\ \hline
12                       & GRU \cite{syu2020ultra}                      & \multicolumn{1}{l|}{0.0228}          & \multicolumn{1}{l|}{0.3797}          & \multicolumn{1}{l|}{0.0038}          & \multicolumn{1}{l|}{6.6361× 10$^{-04}$}   & 0.0258          \\ \hline
13                       & RRBFN \cite{madhiarasan2020accurate}                     & \multicolumn{1}{l|}{0.0111}          & \multicolumn{1}{l|}{0.1845}          & \multicolumn{1}{l|}{0.0018}          & \multicolumn{1}{l|}{3.5161×   10$^{-04}$} & 0.0188          \\ \hline
14                       & PDBN \cite{zhang2015predictive}                     & \multicolumn{1}{l|}{0.0061}          & \multicolumn{1}{l|}{0.1014}          & \multicolumn{1}{l|}{0.0010}          & \multicolumn{1}{l|}{5.8615×   10$^{-05}$} & 0.0077          \\ \hline
15                       & CEEMDAN-SVR \cite{ren2014comparative}               & \multicolumn{1}{l|}{0.0168}          & \multicolumn{1}{l|}{0.2786}          & \multicolumn{1}{l|}{0.0028}          & \multicolumn{1}{l|}{3.6891×   10$^{-04}$} & 0.0192          \\ \hline
16                       & VMD-PSO-LSTM \cite{li2020hybrid}             & \multicolumn{1}{l|}{0.0056}          & \multicolumn{1}{l|}{0.0935}          & \multicolumn{1}{l|}{9.3504×   10$^{-04}$} & \multicolumn{1}{l|}{5.3236×   10$^{-05}$} & 0.0073          \\ \hline
17                       & RBFN-LSSVM \cite{shi2013hybrid}                 & \multicolumn{1}{l|}{0.0154}          & \multicolumn{1}{l|}{0.2552}          & \multicolumn{1}{l|}{0.0026}          & \multicolumn{1}{l|}{2.8649×   10$^{-04}$} & 0.0169          \\ \hline
18                       & VMD-DE-ESN \cite{hu2021wind}               & \multicolumn{1}{l|}{0.0038}          & \multicolumn{1}{l|}{0.0630}          & \multicolumn{1}{l|}{6.3048×   10$^{-04}$} & \multicolumn{1}{l|}{2.4677×   10$^{-05}$} & 0.0050          \\ \hline
19                       & RSDAE \cite{khodayar2017rough}                     & \multicolumn{1}{l|}{0.0238}          & \multicolumn{1}{l|}{0.3949}          & \multicolumn{1}{l|}{0.0039}          & \multicolumn{1}{l|}{7.3605×   10$^{-04}$} & 0.0271          \\ \hline
20                       & MMLSTM \cite{li2020ultra}                    & \multicolumn{1}{l|}{0.0042}          & \multicolumn{1}{l|}{0.0699}          & \multicolumn{1}{l|}{6.9928×   10$^{-04}$} & \multicolumn{1}{l|}{5.3099×   10$^{-05}$} & 0.0073          \\ \hline
21                       & DMDNN \cite{afrasiabi2020advanced}                    & \multicolumn{1}{l|}{0.0027}          & \multicolumn{1}{l|}{0.0442}          & \multicolumn{1}{l|}{4.4200×   10$^{-04}$} & \multicolumn{1}{l|}{1.6218×   10$^{-05}$} & 0.0040          \\ \hline
22                       & VMD-ConvLSTM-ES \cite{sun2020short}           & \multicolumn{1}{l|}{0.0014}          & \multicolumn{1}{l|}{0.0226}          & \multicolumn{1}{l|}{2.2649×   10$^{-04}$} & \multicolumn{1}{l|}{3.6718×   10$^{-06}$} & 0.0019          \\ \hline
23                       & ICEEMDAN-GRU-ARIMA-ECM \cite{duan2021short}    & \multicolumn{1}{l|}{8.3311×   10$^{-04}$} & \multicolumn{1}{l|}{0.0138}          & \multicolumn{1}{l|}{1.3848×   10$^{-04}$} & \multicolumn{1}{l|}{1.5061×   10$^{-06}$} & 0.0012          \\ \hline
24                       & Proposed ICEEMDAN-TNF              & \multicolumn{1}{l|}{4.4077×   10$^{-05}$} & \multicolumn{1}{l|}{7.3264×   10$^{-04}$} & \multicolumn{1}{l|}{7.3264×   10$^{-06}$} & \multicolumn{1}{l|}{1.8512×   10$^{-08}$} & 1.3606×   10$^{-04}$ \\ \hline
25                       &\textbf{Proposed ICEEMDAN-TNF-MLPN-RECS}    & \multicolumn{1}{l|}{\textbf{1.7096× 10$^{-07}$} }  & \multicolumn{1}{l|}{\textbf{2.8416× 10$^{-06}$}} & \multicolumn{1}{l|}{\textbf{2.8416× 10$^{-08}$} }  & \multicolumn{1}{l|}{\textbf{5.0206× 10$^{-14}$}}  & \textbf{2.2407× 10$^{-07}$}   \\ \hline
\end{tabular}
\end{table}

\begin{table}[]
\tiny
\centering
\caption{COMPARATIVE PERFORMANCE RESULTS ANALYSIS FOR CASE STUDY 2. Bold represents the best results.}
\label{Table 4}
\begin{tabular}{|p{0.5cm}|p{1cm}|lllll|}
\hline
\multirow{2}{*}{\textbf{S.   NO}} & \multirow{2}{*}{\textbf{\makecell{Forecasting \\ Model}} }& \multicolumn{5}{l|}{\textbf{Performance Error Index} }                                                                                                                              \\ \cline{3-7} 
                         &                                    & \multicolumn{1}{l|}{MAE}             & \multicolumn{1}{l|}{MAPE}            & \multicolumn{1}{l|}{MRE}             & \multicolumn{1}{l|}{MSE}             & RMSE            \\ \hline
1                        & Persistent \cite{franke1995investigations}                & \multicolumn{1}{l|}{6.1802}          & \multicolumn{1}{l|}{95.3705}         & \multicolumn{1}{l|}{0.9537}          & \multicolumn{1}{l|}{46.9371}         & 6.851           \\ \hline
2                        & NWP \cite{landberg1994short}                        & \multicolumn{1}{l|}{1.6664}          & \multicolumn{1}{l|}{25.7153}         & \multicolumn{1}{l|}{0.2572}          & \multicolumn{1}{l|}{0.6609}          & 2.5809          \\ \hline
3                        & ARIMA \cite{palomares2009arima}                     & \multicolumn{1}{l|}{0.8664}          & \multicolumn{1}{l|}{13.3701}         & \multicolumn{1}{l|}{0.1337}          & \multicolumn{1}{l|}{3.0771}          & 1.7542          \\ \hline
4                        & BPNN \cite{madhiarasan2016performance}                      & \multicolumn{1}{l|}{0.7619}          & \multicolumn{1}{l|}{11.7581}         & \multicolumn{1}{l|}{0.1176}          & \multicolumn{1}{l|}{5.1488}          & 2.2691          \\ \hline
5                        & ENN \cite{madhiarasan2016performance}                       & \multicolumn{1}{l|}{0.5251}          & \multicolumn{1}{l|}{8.1050}          & \multicolumn{1}{l|}{0.0811}          & \multicolumn{1}{l|}{1.3950}          & 1.1811          \\ \hline
6                        & RNN \cite{barbounis2006long}                        & \multicolumn{1}{l|}{0.4415}          & \multicolumn{1}{l|}{6.8130}          & \multicolumn{1}{l|}{0.0681}          & \multicolumn{1}{l|}{0.7463}          & 0.8639          \\ \hline
7                        & RBFN \cite{junli2010wind}                       & \multicolumn{1}{l|}{0.2512}          & \multicolumn{1}{l|}{3.8768}          & \multicolumn{1}{l|}{0.0388}          & \multicolumn{1}{l|}{0.1228}          & 0.3505          \\ \hline
8                        & SVR \cite{botha2017forecasting}                       & \multicolumn{1}{l|}{0.2503}          & \multicolumn{1}{l|}{3.8624}          & \multicolumn{1}{l|}{0.0386}          & \multicolumn{1}{l|}{0.4615}          & 0.6793          \\ \hline
9                        & XGBoost \cite{ahmadi2020long}                   & \multicolumn{1}{l|}{0.1502}          & \multicolumn{1}{l|}{2.3181}          & \multicolumn{1}{l|}{0.0232}          & \multicolumn{1}{l|}{0.0548}          & 0.2340          \\ \hline
10                       & MLPN \cite{madhiarasan2017comparative}                      & \multicolumn{1}{l|}{0.0421}          & \multicolumn{1}{l|}{0.6493}          & \multicolumn{1}{l|}{0..0065}         & \multicolumn{1}{l|}{0.0303}          & 0.1740          \\ \hline
11                       & LSTM \cite{araya2018lstm}                      & \multicolumn{1}{l|}{0.0321}          & \multicolumn{1}{l|}{0.4952}          & \multicolumn{1}{l|}{0.0050}          & \multicolumn{1}{l|}{0.0020}          & 0.0448          \\ \hline
12                       & GRU \cite{syu2020ultra}                      & \multicolumn{1}{l|}{0.0258}          & \multicolumn{1}{l|}{0.3977}          & \multicolumn{1}{l|}{0.0040}          & \multicolumn{1}{l|}{0.0019}          & 0.0439          \\ \hline
13                       & RRBFN \cite{madhiarasan2020accurate}                     & \multicolumn{1}{l|}{0.0115}          & \multicolumn{1}{l|}{0.1774}          & \multicolumn{1}{l|}{0.0018}          & \multicolumn{1}{l|}{1.8105×   10$^{-04}$} & 0.0135          \\ \hline
14                       & PDBN \cite{zhang2015predictive}                     & \multicolumn{1}{l|}{0.0074}          & \multicolumn{1}{l|}{0.1135}          & \multicolumn{1}{l|}{0.0011}          & \multicolumn{1}{l|}{7.5029×   10$^{-05}$} & 0.0087          \\ \hline
15                       & CEEMDAN-SVR \cite{ren2014comparative}               & \multicolumn{1}{l|}{0.0163}          & \multicolumn{1}{l|}{0.2521}          & \multicolumn{1}{l|}{0.0025}          & \multicolumn{1}{l|}{3.3222×   10$^{-04}$} & 0.0182          \\ \hline
16                       & VMD-PSO-LSTM \cite{li2020hybrid}              & \multicolumn{1}{l|}{0.0060}          & \multicolumn{1}{l|}{0.0930}          & \multicolumn{1}{l|}{9.3036×   10$^{-04}$} & \multicolumn{1}{l|}{5.6657×   10$^{-05}$} & 0.0075          \\ \hline
17                       & RBFN-LSSVM \cite{shi2013hybrid}                & \multicolumn{1}{l|}{0.0161}          & \multicolumn{1}{l|}{0.2485}          & \multicolumn{1}{l|}{0.0025}          & \multicolumn{1}{l|}{3.4529×   10$^{-04}$} & 0.0186          \\ \hline
18                       & VMD-DE-ESN \cite{hu2021wind}               & \multicolumn{1}{l|}{0.0043}          & \multicolumn{1}{l|}{0.0656}          & \multicolumn{1}{l|}{6.5645×   10$^{-04}$} & \multicolumn{1}{l|}{4.2924×   10$^{-05}$} & 0.0066          \\ \hline
19                       & RSDAE \cite{khodayar2017rough}                     & \multicolumn{1}{l|}{0.0272}          & \multicolumn{1}{l|}{0.4202}          & \multicolumn{1}{l|}{0.0042}          & \multicolumn{1}{l|}{0.0010}          & 0.0322          \\ \hline
20                       & MMLSTM \cite{li2020ultra}                    & \multicolumn{1}{l|}{0.0054}          & \multicolumn{1}{l|}{0.0827}          & \multicolumn{1}{l|}{8.2685×   10$^{-05}$} & \multicolumn{1}{l|}{4.7116×   10$^{-05}$} & 0.0069          \\ \hline
21                       & DMDNN \cite{afrasiabi2020advanced}                     & \multicolumn{1}{l|}{0.0032}          & \multicolumn{1}{l|}{0.0496}          & \multicolumn{1}{l|}{4.9607×   10$^{-04}$} & \multicolumn{1}{l|}{2.1454×   10$^{-05}$} & 0.0046          \\ \hline
22                       & VMD-ConvLSTM-ES \cite{sun2020short}          & \multicolumn{1}{l|}{0.0018}          & \multicolumn{1}{l|}{0.0280}          & \multicolumn{1}{l|}{2.8007×   10$^{-04}$} & \multicolumn{1}{l|}{7.8080×   10$^{-06}$} & 0.0028          \\ \hline
23                       & ICEEMDAN-GRU-ARIMA-ECM \cite{duan2021short}   & \multicolumn{1}{l|}{7.5599×   10$^{-04}$} & \multicolumn{1}{l|}{0.0117}          & \multicolumn{1}{l|}{1.1666×   10$^{-04}$} & \multicolumn{1}{l|}{1.2393×   10$^{-06}$} & 0.0011          \\ \hline
24                       & Proposed ICEEMDAN-TNF              & \multicolumn{1}{l|}{5.0588×   10$^{-05}$} & \multicolumn{1}{l|}{7.8065×   10$^{-04}$} & \multicolumn{1}{l|}{7.8065×   10$^{-06}$} & \multicolumn{1}{l|}{5.2334×   10$^{-07}$} & 7.2343×   10$^{-04}$ \\ \hline
25                       & \textbf{Proposed ICEEMDAN-TNF-MLPN-RECS}   & \multicolumn{1}{l|}{\textbf{6.1565× 10$^{-07}$}}  & \multicolumn{1}{l|}{\textbf{9.5005× 10$^{-06}$}}  & \multicolumn{1}{l|}{\textbf{9.5005× 10$^{-08}$}}   & \multicolumn{1}{l|}{\textbf{8.9289× 10$^{-13}$}}   & \textbf{9.4493× 10$^{-07}$} \\ \hline
\end{tabular}
\end{table}
\subsection{Performance Analysis for Case Study 2}
A comparative investigation was performed on the case study 2 wind farm dataset, and the forecasting accuracy was evaluated using the performance error index. The proposed hybrid model-based results have outperformed wind speed forecasting compared to the state-of-the-art and baseline forecasting models. Case study 2: the performance error index based on the comparative performance analysis plot is shown in Figure \ref{Figure 6} (b) for better understanding. 
Each method-based forecasting is realized using the real-time wind farm data and computes the performance error index. We derived the following discussion regarding both case studies based on performance investigation: The persistent model error index is comparatively much higher than other models. ARIMA model performance is better than NWP, but less than BPNN. Elman neural network error indexes are less than BPNN, ARIMA, NWP but greater than RNN. XGBoost error values are less than SVR and RBFN. GRU performance is slightly better than MLPN, LSTM, but RRBFN performs better than GRU and less than PDB. The PDBN error indexes are less than CEEMDAN-SVR, RBFN-LSSVM, and RSDAE, but VMD-PSO-LSTM error-indexes are less than PDBN. DMDNN performance is better than MMLSTM, VMD-DE-ESN, VMD-DE-ESN, and VMD-PSO-LSTM but less VMD-ConvLSTM-ES. The ICEEMDAN-GRU-ARIMA-ECM model performance error index is less than other models. Still, our proposed model surpassed the state-of-the-art model and outperformed better than all comparatively analyzed models in both case studies, which confirms that the proposed model excelled in baseline and state-of-the-art forecasting models.
The underlying summary from the comparative analysis is the proposed hybrid forecasting model significantly minimizes the performance error index (MAE, MAPE, MRE, MSE, and RMSE), improving the forecasting exactness. 
\par From the comparative analysis, arrives the conclusion that the proposed model enhances the wind speed forecasting quality. Our experimental simulation outputs suggest that adding error correction further enriches the forecasting model performance. Figure \ref{Figure 6} illustrate the high-level correct forecasting capability of the proposed model with a minimal error index compared to other forecasting models. The accuracy and sharpness of the proposed hybrid forecasting model are validated using two hub height-based collected real-time wind farm data from the location Kethanur, India. Compared with the other models, the proposed model is surpassed, which is inferred from Tables \ref{Table 3} and \ref{Table 4}. 
\par The simulation-based numerical results compared with the current state-of-the-method and classical baseline methods, the performance analysis of the proposed ICEEMDAN-TNF without RECS, and ICEEMDAN-TNF-MLPN-RECS significantly minimize the performance error index to the very minimal value than other models. A complex and dynamic presence in the wind farm data is effectively learned with the help of the transformer network and ICEEMDAN, which favorably makes it superior to the other state-of-the-art methods used for the comparative analysis. The performance analysis infers that the leverage of adding the residual error correction strategy into the hybrid model is precise wind speed forecasting.


\section{Different Horizons Wind Speed Forecasting}
We can classify wind speed forecasting based on the time horizons as long-term, medium-term, short-term and very short-term horizons. Timely need for different horizons based on consistent and robust forecasting models for adequate wind energy integrated smart grid. We analyzed the proposed ICEEMDAN-TNF-MLPN-RECS model performance validity regarding different horizons based on wind speed forecasting for both case studies.
\begin {figure}
\centering
\includegraphics [width=0.5\textwidth]{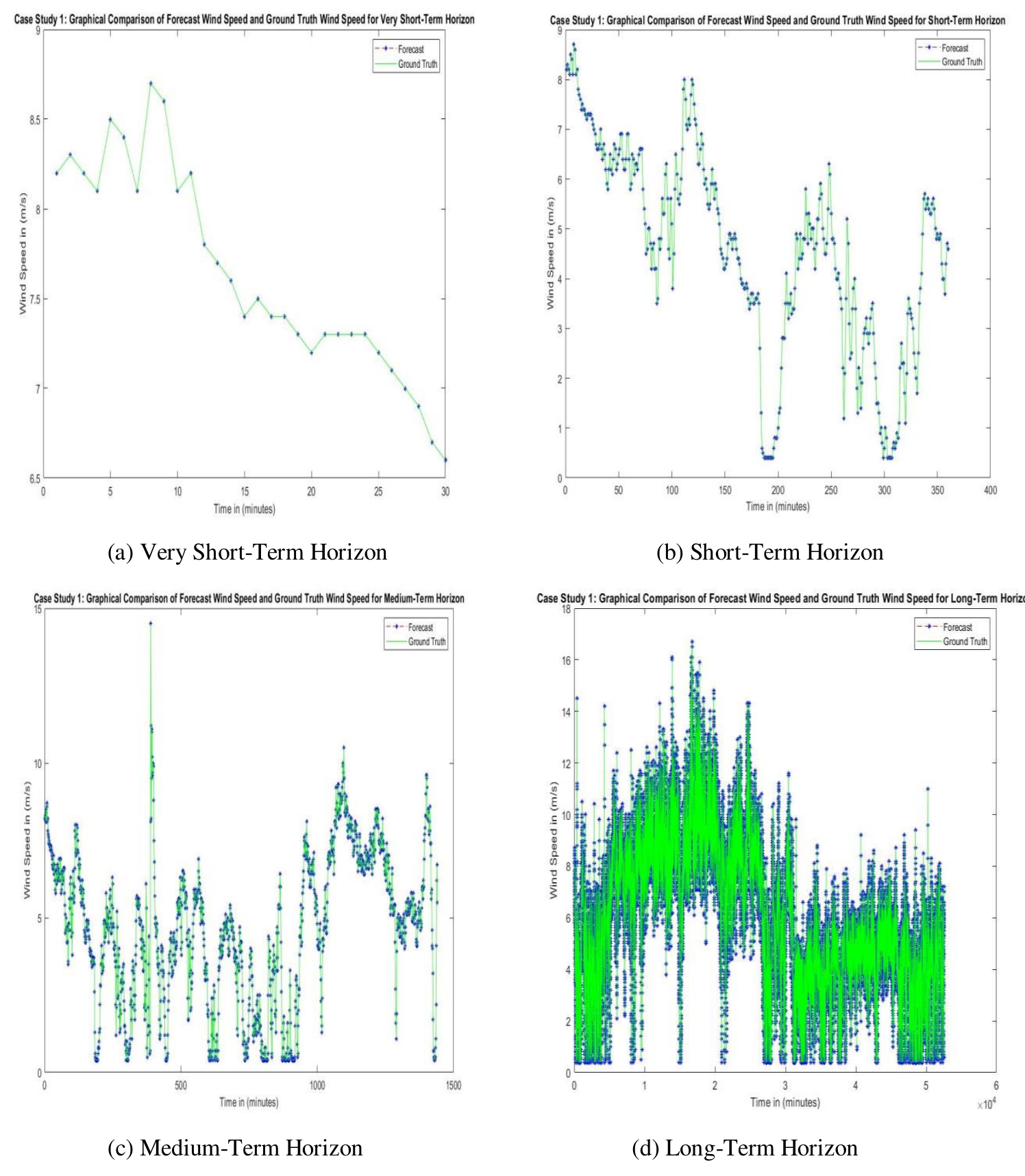}
\caption{Case Study 1: Proposed Different Horizon Forecasting Models based on Comparison of Forecast and Ground Truth Wind Speed.}
\centering
\label{Figure 7}
\end{figure}
\begin {figure}
\centering
\includegraphics [width=0.5\textwidth]{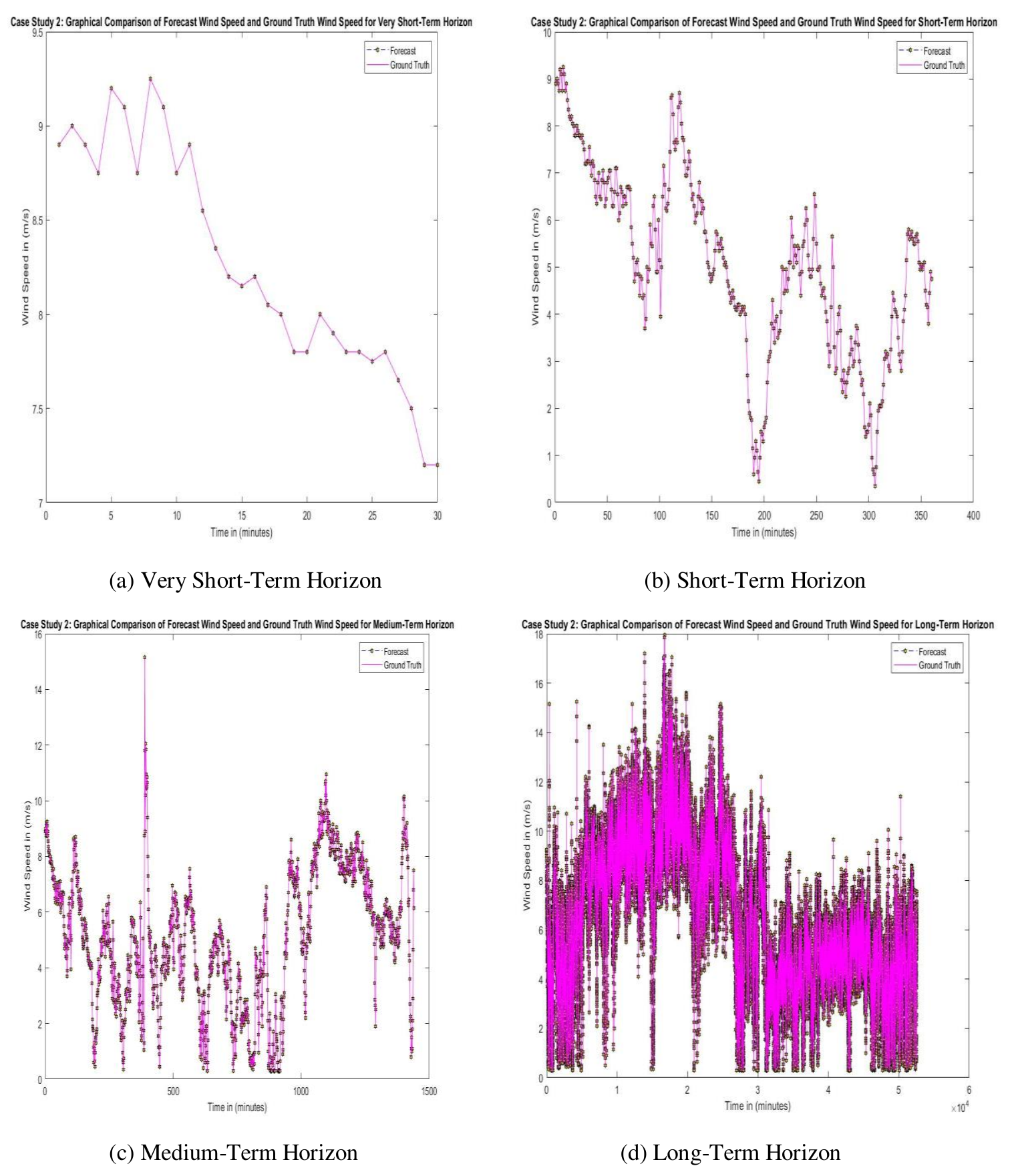}
\caption{Case Study 2: Proposed Different Horizon Forecasting Models based on Comparison of Forecast and Ground Truth Wind Speed.}
\centering
\label{Figure 8}
\end{figure}
\subsection{Different Horizons Forecasting for Case Study 1}
Wind speed forecasting on different horizons based on performance analyses of the proposed ICEEMDAN-TNF-MLPN-RECS Model for case study 1 performed and achieved experimental simulation results are tabulated in Table \ref{Table 5}. Stabilized outputs are guaranteed on different horizons based on wind speed forecasting by the proposed hybrid model (ICEEMDAN-TNF-MLPN-RECS). Figure \ref{Figure 7} (a), (b), (c) and (d) display the obtained results of the proposed ICEEMDAN-TNF-MLPN-RECS model based on wind speed forecasting for very short-term, short-term, medium-term, and long-term, respectively, for the case study 1. Although the proposed ICEEMDAN-TNF-MLPN-RECS model achieved promising wind speed forecasting with different horizons, the proposed model-based very short-term wind speed forecasting holds the lowest performance error index such as MAE-4.5729× 10$^{-08}$, MAPE-5.9803× 10$^{-07}$, MRE-5.9803× 10$^{-09}$, MSE-2.2555× 10$^{-14}$, RMSE-1.5018× 10$^{-07}$. The proposed model is based on very short wind speed forecasting performs better than the other forecasting horizons. The medium-term horizon-based forecasting accuracy is higher than the long-term and short-term horizons. Similarly, long-term horizon-based wind speed forecasting achieves better accuracy than the short-term horizon. 
\begin{table}[]
\tiny
\caption{CASE STUDY 1: DIFFERENT HORIZON-BASED EXPERIMENTAL RESULTS.}
\label{Table 5}
\begin{tabular}{|l|lllll|}
\hline
\multirow{2}{*}{\textbf{Wind Speed Forecast Horizons}} &
  \multicolumn{5}{l|}{\textbf{Performance Error Index} }\\ \cline{2-6} 
 &
  \multicolumn{1}{l|}{\textbf{MAE}} &
  \multicolumn{1}{l|}{\textbf{MAPE}} &
  \multicolumn{1}{l|}{\textbf{MRE}} &
  \multicolumn{1}{l|}{\textbf{MSE}} &
  \textbf {RMSE} \\ \hline
Very Short-Term Horizon &
  \multicolumn{1}{l|}{4.5729× 10$^{-08}$} &
  \multicolumn{1}{l|}{5.9803× 10$^{-07}$} &
  \multicolumn{1}{l|}{5.9803× 10$^{-09}$} &
  \multicolumn{1}{l|}{2.2555× 10$^{-14}$} &
  1.5018× 10$^{-07}$ \\ \hline
Short-Term Horizon &
  \multicolumn{1}{l|}{2.8093 × 10$^{-07}$} &
  \multicolumn{1}{l|}{6.1893× 10$^{-06}$} &
  \multicolumn{1}{l|}{6.1893× 10$^{-08}$} &
  \multicolumn{1}{l|}{4.4394× 10$^{-12}$} &
  2.1070× 10$^{-06}$ \\ \hline
Medium-Term Horizon &
  \multicolumn{1}{l|}{1.4146× 10$^{-07}$} &
  \multicolumn{1}{l|}{3.0799× 10$^{-06}$} &
  \multicolumn{1}{l|}{3.0799× 10$^{-08}$} &
  \multicolumn{1}{l|}{2.8908× 10$^{-14}$} &
  1.7002× 10$^{-07}$ \\ \hline
Long-Term Horizon &
  \multicolumn{1}{l|}{1.7096× 10$^{-07}$} &
  \multicolumn{1}{l|}{2.8416× 10$^{-06}$} &
  \multicolumn{1}{l|}{2.8416× 10$^{-08}$} &
  \multicolumn{1}{l|}{5.0206× 10$^{-14}$} &
  2.2407× 10$^{-07}$ \\ \hline
\end{tabular}
\end{table}

\subsection{Different Horizons Forecasting for Case Study 2}
The proposed ICEEMDAN-TNF-MLPN-RECS Model performance is analyzed for case study 2 regarding different horizons based on wind speed forecasting and the obtained experimental simulation results are tabulated in Table \ref{Table 6}. Figure \ref{Figure 8} (a), (b), (c) and (d) display the obtained results of the proposed ICEEMDAN-TNF-MLPN-RECS model based on wind speed forecasting for very short-term, short-term, medium-term, and long-term, respectively, for the case study 2. Although the proposed ICEEMDAN-TNF-MLPN-RECS model performs well on different horizons based on wind speed forecasting, the very short-term wind speed forecasting retains the lowest performance error index such as MAE-4.3359× 10$^{-07}$, MAPE-5.2387× 10$^{-06}$, MRE-5.2387× 10$^{-08}$, MSE-5.8036× 10$^{-13}$, RMSE-7.6181× 10$^{-07}$ than other forecasting horizons. The medium-term horizon-based forecasting performance error indexes are less than the long-term and short-term horizons. Similarly, long-term horizon-based wind speed forecasting achieves less performance error index than the short-term horizon. \\
\begin{table}[]
\tiny
\caption{CASE STUDY 2: DIFFERENT HORIZON-BASED EXPERIMENTAL RESULTS.}
\label{Table 6}
\begin{tabular}{|l|lllll|}
\hline
\multirow{2}{*}{\textbf{Wind Speed Forecast Horizons}} &
  \multicolumn{5}{l|}{\textbf{Performance Error Index} }\\ \cline{2-6} 
 &
  \multicolumn{1}{l|}{\textbf{MAE}} &
  \multicolumn{1}{l|}{\textbf{MAPE}} &
  \multicolumn{1}{l|}{\textbf{MRE}} &
  \multicolumn{1}{l|}{\textbf{MSE}} &
  \textbf {RMSE} \\ \hline
Very Short-Term Horizon &
  \multicolumn{1}{l|}{4.3359× 10$^{-07}$} &
  \multicolumn{1}{l|}{5.2387× 10$^{-06}$} &
  \multicolumn{1}{l|}{5.2387× 10$^{-08}$} &
  \multicolumn{1}{l|}{5.8036× 10$^{-13}$} &
  7.6181× 10$^{-07}$ \\ \hline
Short-Term Horizon &
  \multicolumn{1}{l|}{1.1718× 10$^{-06}$} &
  \multicolumn{1}{l|}{2.3351× 10$^{-05}$} &
  \multicolumn{1}{l|}{2.3351× 10$^{-07}$} &
  \multicolumn{1}{l|}{1.4710× 10$^{-11}$} &
  3.8354× 10$^{-06}$ \\ \hline
Medium-Term Horizon &
  \multicolumn{1}{l|}{5.1609× 10$^{-07}$} &
  \multicolumn{1}{l|}{1.0139× 10$^{-05}$} &
  \multicolumn{1}{l|}{1.0139× 10$^{-07}$} &
  \multicolumn{1}{l|}{4.2912× 10$^{-13}$} &
  6.5507× 10$^{-07}$ \\ \hline
Long-Term Horizon &
  \multicolumn{1}{l|}{6.1565× 10$^{-07}$} &
  \multicolumn{1}{l|}{9.5005× 10$^{-06}$} &
  \multicolumn{1}{l|}{9.5005× 10$^{-08}$} &
  \multicolumn{1}{l|}{8.9289× 10$^{-13}$} &
  9.4493× 10$^{-07}$ \\ \hline
\end{tabular}
\end{table}
Tables \ref{Table 5} and \ref{Table 6} show that the proposed hybrid model is suitable to forecast the different horizons based on wind speed with high accuracy and results in the lowest performance error indexes. 
Two case studies analyzed the proposed ICEEMDAN-TNF-MLPN-RECS forecasting model, predominating wind speed forecasting for different horizons. We explicitly understood the success of the proposed model-based wind speed forecast for different horizons from the empirical results.
\section{Conclusion}
The proposed ICEEMDAN-TNF-MLPN-RECS hybrid model forecasting capability is validated on the real-time two wind turbine hub-based retrieved wind farm data. We carried out a comparative performance investigation to demonstrate the effectiveness of the proposed hybrid (ICEEMDAN-TNF-MLPN-RECS) wind speed forecasting model. According to the comparative investigation Tables \ref{Table 3} and \ref{Table 4}, it is realized that the proposed wind speed forecasting model is outperforming with a high degree of precise forecasting and generalizing to both case studies. Moreover, we overcome the highlighted general issues of the baseline and other forecasting models. In addition, we performed different horizon-based wind speed forecasting to confirm the validity of the proposed hybrid (ICEEMDAN-TNF-MLPN-RECS) with different horizons based on wind speed forecasting. Thus, the proposed forecasting is capable enough to support the modern power system operator, effectively balancing demand and generation. The proposed intelligent hybrid forecasting model is generic and applicable to other forecasting application scenarios. Future research plans to extend the work to various time scales, with multivariate-based wind speed forecasting and real-time practical implementation of the wind farm.
\small
\bibliography{IEEEabrv,bibtex/bib/paper}

\begin{thebibliography}{29}
\providecommand{\natexlab}[1]{#1}
\providecommand{\url}[1]{#1}
\csname url@samestyle\endcsname
\providecommand{\newblock}{\relax}
\providecommand{\bibinfo}[2]{#2}
\providecommand{\BIBentrySTDinterwordspacing}{\spaceskip=0pt\relax}
\providecommand{\BIBentryALTinterwordstretchfactor}{4}
\providecommand{\BIBentryALTinterwordspacing}{\spaceskip=\fontdimen2\font plus
\BIBentryALTinterwordstretchfactor\fontdimen3\font minus
  \fontdimen4\font\relax}
\providecommand{\BIBforeignlanguage}[2]{{%
\expandafter\ifx\csname l@#1\endcsname\relax
\typeout{** WARNING: IEEEtranN.bst: No hyphenation pattern has been}%
\typeout{** loaded for the language `#1'. Using the pattern for}%
\typeout{** the default language instead.}%
\else
\language=\csname l@#1\endcsname
\fi
#2}}
\providecommand{\BIBdecl}{\relax}
\BIBdecl

\bibitem[Madhiarasan(2018)]{madhiarasancertain}
M.~Madhiarasan, ``Certain algebraic criteria for design of hybrid neural
  network models with applications in renewable energy forecasting,'' Ph.D.
  thesis, Faculty of Electrical Engineering, Anna University, 2018.

\bibitem[Madhiarasan and Deepa(2016{\natexlab{a}})]{madhiarasan2016novel}
M.~Madhiarasan and S.~Deepa, ``A novel criterion to select hidden neuron
  numbers in improved back propagation networks for wind speed forecasting,''
  \emph{Applied intelligence}, vol.~44, no.~4, pp. 878--893, 2016.

\bibitem[Franke(1995)]{franke1995investigations}
J.~Franke, ``Investigations into the dynamics of wakes in the atmospheric
  boundary layer; untersuchungen zur dynamik von wirbelschleppen in der
  atmosphaerischen grenzschicht,'' Ph.D. thesis, Berichte des Instituts für
  Meteorologie und Klimatologie der Universität Hannover (in German), 1995.

\bibitem[Landberg and Watson(1994)]{landberg1994short}
L.~Landberg and S.~J. Watson, ``Short-term prediction of local wind
  conditions,'' \emph{Boundary-Layer Meteorology}, vol.~70, no.~1, pp.
  171--195, 1994.

\bibitem[Palomares-Salas et~al.(2009)Palomares-Salas, De~La~Rosa, Ramiro,
  Melgar, Aguera, and Moreno]{palomares2009arima}
J.~Palomares-Salas, J.~De~La~Rosa, J.~Ramiro, J.~Melgar, A.~Aguera, and
  A.~Moreno, ``Arima vs. neural networks for wind speed forecasting,'' in
  \emph{2009 IEEE International Conference on Computational Intelligence for
  Measurement Systems and Applications}.\hskip 1em plus 0.5em minus 0.4em\relax
  IEEE, 2009, pp. 129--133.

\bibitem[Madhiarasan and Deepa(2016{\natexlab{b}})]{madhiarasan2016performance}
M.~Madhiarasan and S.~Deepa, ``Performance investigation of six artificial
  neural networks for different time scale wind speed forecasting in three wind
  farms of coimbatore region,'' \emph{International Journal of Innovation and
  Scientific Research}, vol.~23, no.~2, pp. 380--411, 2016.

\bibitem[Barbounis et~al.(2006)Barbounis, Theocharis, Alexiadis, and
  Dokopoulos]{barbounis2006long}
T.~G. Barbounis, J.~B. Theocharis, M.~C. Alexiadis, and P.~S. Dokopoulos,
  ``Long-term wind speed and power forecasting using local recurrent neural
  network models,'' \emph{IEEE Transactions on Energy Conversion}, vol.~21,
  no.~1, pp. 273--284, 2006.

\bibitem[Junli et~al.(2010)Junli, Xingjie, and Jian]{junli2010wind}
W.~Junli, L.~Xingjie, and Q.~Jian, ``Wind speed and power forecasting based on
  rbf neural network,'' in \emph{2010 International Conference on Computer
  Application and System Modeling (ICCASM 2010)}, vol.~5.\hskip 1em plus 0.5em
  minus 0.4em\relax IEEE, 2010, pp. V5--298.

\bibitem[Botha and van~der Walt(2017)]{botha2017forecasting}
N.~Botha and C.~M. van~der Walt, ``Forecasting wind speed using support vector
  regression and feature selection,'' in \emph{2017 Pattern Recognition
  Association of South Africa and Robotics and Mechatronics
  (PRASA-RobMech)}.\hskip 1em plus 0.5em minus 0.4em\relax IEEE, 2017, pp.
  181--186.

\bibitem[Ahmadi et~al.(2020)Ahmadi, Nabipour, Mohammadi-Ivatloo, Amani, Rho,
  and Piran]{ahmadi2020long}
A.~Ahmadi, M.~Nabipour, B.~Mohammadi-Ivatloo, A.~M. Amani, S.~Rho, and M.~J.
  Piran, ``Long-term wind power forecasting using tree-based learning
  algorithms,'' \emph{IEEE Access}, vol.~8, pp. 151\,511--151\,522, 2020.

\bibitem[Madhiarasan and Deepa(2017)]{madhiarasan2017comparative}
M.~Madhiarasan and S.~Deepa, ``Comparative analysis on hidden neurons
  estimation in multi layer perceptron neural networks for wind speed
  forecasting,'' \emph{Artificial Intelligence Review}, vol.~48, no.~4, pp.
  449--471, 2017.

\bibitem[Araya et~al.(2018)Araya, Valle, and Allende]{araya2018lstm}
I.~A. Araya, C.~Valle, and H.~Allende, ``Lstm-based multi-scale model for wind
  speed forecasting,'' in \emph{Iberoamerican Congress on Pattern
  Recognition}.\hskip 1em plus 0.5em minus 0.4em\relax Springer, 2018, pp.
  38--45.

\bibitem[Syu et~al.(2020)Syu, Wang, Chou, Lin, Liang, Wu, and
  Jiang]{syu2020ultra}
Y.-D. Syu, J.-C. Wang, C.-Y. Chou, M.-J. Lin, W.-C. Liang, L.-C. Wu, and J.-A.
  Jiang, ``Ultra-short-term wind speed forecasting for wind power based on
  gated recurrent unit,'' in \emph{2020 8th International Electrical
  Engineering Congress (iEECON)}.\hskip 1em plus 0.5em minus 0.4em\relax IEEE,
  2020, pp. 1--4.

\bibitem[Madhiarasan(2020)]{madhiarasan2020accurate}
M.~Madhiarasan, ``Accurate prediction of different forecast horizons wind speed
  using a recursive radial basis function neural network,'' \emph{Protection
  and Control of Modern Power Systems}, vol.~5, no.~1, pp. 1--9, 2020.

\bibitem[Zhang et~al.(2015)Zhang, Chen, Gan, and Chen]{zhang2015predictive}
C.-Y. Zhang, C.~P. Chen, M.~Gan, and L.~Chen, ``Predictive deep boltzmann
  machine for multiperiod wind speed forecasting,'' \emph{IEEE Transactions on
  Sustainable Energy}, vol.~6, no.~4, pp. 1416--1425, 2015.

\bibitem[Ren et~al.(2014)Ren, Suganthan, and Srikanth]{ren2014comparative}
Y.~Ren, P.~Suganthan, and N.~Srikanth, ``A comparative study of empirical mode
  decomposition-based short-term wind speed forecasting methods,'' \emph{IEEE
  Transactions on Sustainable Energy}, vol.~6, no.~1, pp. 236--244, 2014.

\bibitem[Li et~al.(2020{\natexlab{a}})Li, Chen, Li, Tang, Gan, and
  An]{li2020hybrid}
Y.~Li, X.~Chen, C.~Li, G.~Tang, Z.~Gan, and X.~An, ``A hybrid deep interval
  prediction model for wind speed forecasting,'' \emph{IEEE Access}, vol.~9,
  pp. 7323--7335, 2020.

\bibitem[Shi et~al.(2013)Shi, Ding, Lee, Yang, Liu, and Zhang]{shi2013hybrid}
J.~Shi, Z.~Ding, W.-J. Lee, Y.~Yang, Y.~Liu, and M.~Zhang, ``Hybrid forecasting
  model for very-short term wind power forecasting based on grey relational
  analysis and wind speed distribution features,'' \emph{IEEE Transactions on
  Smart Grid}, vol.~5, no.~1, pp. 521--526, 2013.

\bibitem[Hu et~al.(2021)Hu, Wang, and Tao]{hu2021wind}
H.~Hu, L.~Wang, and R.~Tao, ``Wind speed forecasting based on variational mode
  decomposition and improved echo state network,'' \emph{Renewable Energy},
  vol. 164, pp. 729--751, 2021.

\bibitem[Khodayar et~al.(2017)Khodayar, Kaynak, and
  Khodayar]{khodayar2017rough}
M.~Khodayar, O.~Kaynak, and M.~E. Khodayar, ``Rough deep neural architecture
  for short-term wind speed forecasting,'' \emph{IEEE Transactions on
  Industrial Informatics}, vol.~13, no.~6, pp. 2770--2779, 2017.

\bibitem[Zheng et~al.(2021)Zheng, Wang, Yang, and Zhang]{zheng2021generative}
Z.~Zheng, L.~Wang, L.~Yang, and Z.~Zhang, ``Generative probabilistic wind speed
  forecasting: A variational recurrent autoencoder based method,'' \emph{IEEE
  Transactions on Power Systems}, 2021.

\bibitem[Li et~al.(2020{\natexlab{b}})Li, Zhang, Ji, and Wu]{li2020ultra}
M.~Li, Z.~Zhang, T.~Ji, and Q.~Wu, ``Ultra-short term wind speed prediction
  using mathematical morphology decomposition and long short-term memory,''
  \emph{CSEE Journal of Power and Energy Systems}, vol.~6, no.~4, pp. 890--900,
  2020.

\bibitem[Afrasiabi et~al.(2020)Afrasiabi, Mohammadi, Rastegar, and
  Afrasiabi]{afrasiabi2020advanced}
M.~Afrasiabi, M.~Mohammadi, M.~Rastegar, and S.~Afrasiabi, ``Advanced deep
  learning approach for probabilistic wind speed forecasting,'' \emph{IEEE
  Transactions on Industrial Informatics}, vol.~17, no.~1, pp. 720--727, 2020.

\bibitem[Sun and Zhao(2020)]{sun2020short}
Z.~Sun and M.~Zhao, ``Short-term wind power forecasting based on vmd
  decomposition, convlstm networks and error analysis,'' \emph{IEEE Access},
  vol.~8, pp. 134\,422--134\,434, 2020.

\bibitem[Duan et~al.(2021)Duan, Zuo, Bai, Duan, Chang, and Chen]{duan2021short}
J.~Duan, H.~Zuo, Y.~Bai, J.~Duan, M.~Chang, and B.~Chen, ``Short-term wind
  speed forecasting using recurrent neural networks with error correction,''
  \emph{Energy}, vol. 217, p. 119397, 2021.

\bibitem[Wu et~al.(2020)Wu, Green, Ben, and O'Banion]{wu2020deep}
N.~Wu, B.~Green, X.~Ben, and S.~O'Banion, ``Deep transformer models for time
  series forecasting: The influenza prevalence case,'' \emph{arXiv preprint
  arXiv:2001.08317}, 2020.

\bibitem[Wilks(2011)]{wilks2011statistical}
D.~S. Wilks, \emph{Statistical methods in the atmospheric sciences}.\hskip 1em
  plus 0.5em minus 0.4em\relax Academic press, 2011, vol. 100.

\bibitem[Colominas et~al.(2014)Colominas, Schlotthauer, and
  Torres]{colominas2014improved}
M.~A. Colominas, G.~Schlotthauer, and M.~E. Torres, ``Improved complete
  ensemble emd: A suitable tool for biomedical signal processing,''
  \emph{Biomedical Signal Processing and Control}, vol.~14, pp. 19--29, 2014.

\bibitem[Vaswani et~al.(2017)Vaswani, Shazeer, Parmar, Uszkoreit, Jones, Gomez,
  Kaiser, and Polosukhin]{vaswani2017attention}
A.~Vaswani, N.~Shazeer, N.~Parmar, J.~Uszkoreit, L.~Jones, A.~N. Gomez,
  {\L}.~Kaiser, and I.~Polosukhin, ``Attention is all you need,'' in
  \emph{Advances in neural information processing systems}, 2017, pp.
  5998--6008.

\end{thebibliography}
\newpage
\section*{Authors Biography}

%
\vspace{-6.5cm}
\begin{IEEEbiography}[{\includegraphics[width=1in,height=1.25in,clip,keepaspectratio]{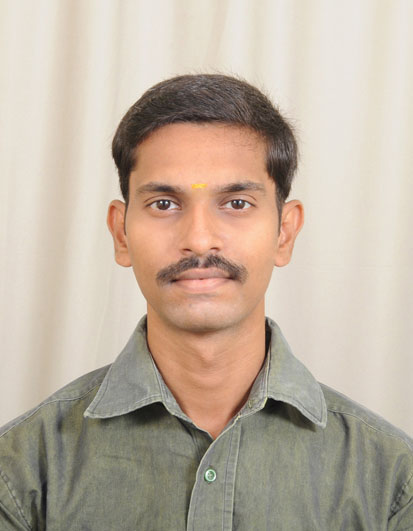}}] {Dr. M. MADHIARASAN} (Member, IEEE) has completed his Bachelor of Engineering degree in Electrical and Electronics Engineering in the year 2010 from Jaya Engineering College, Thiruninravur, under Anna University, Tamil Nadu, India, Master of Engineering degree in Electrical Drives and Embedded Control (Electrical Engineering) in the year 2013 from Anna University, Regional Centre, Coimbatore, under Anna University, Tamil Nadu, India and Ph.D. (Electrical Engineering) in the year 2018 from Anna University, Tamil Nadu, India. \\
He has worked as an Assistant Professor and R \& D In-charge in the Department of Electrical and Electronics Engineering, Bharat Institute of Engineering and Technology, Hyderabad, India, from 2018 to 2020. He is presently working as a Post-Doctoral Fellow at the Department of Computer Science and Engineering, Indian Institute of Technology Roorkee (IITR), India. \\
His research areas include Renewable Energy Systems, Power Electronics and Control, Computer Vision, Human-Computer Interface, Pattern Recognition, Artificial Intelligence, Neural Networks, Optimization, Machine Learning, Deep Learning, Soft Computing, Internet of Things and Modeling and Simulation. \\
He has been a Technical Program Committee Member, International Scientific Committee Member, and Keynote Speaker of many international conferences. He served as Publication Chairs, International Conference on Artificial Intelligence, Big Data and Mechatronics (AIBDM2021). He acts as an editorial board member and reviewer for many peer-reviewed international journals (Springer, Elsevier, IEEE Access, etc.) and a guest editor of Energy Engineering.
\end{IEEEbiography}
\vspace{-5.2cm}
\begin{IEEEbiography}[{\includegraphics[width=1in,height=1.25in,clip,keepaspectratio]{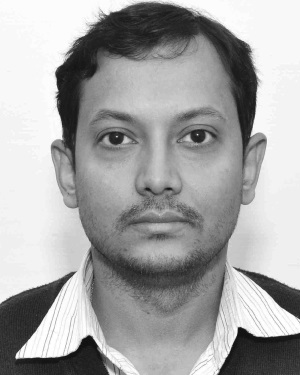}}]{Dr. PARTHA PRATIM ROY} (Member, IEEE) was a Postdoctoral Research Fellow with the RFAI Laboratory, France, in 2012, and the Synchromedia Laboratory, Canada, in 2013. He was with the Advanced Technology Group, Samsung Research Institute, Noida, India, from 2013 to 2014. He is currently an Associate Professor with the Department of Computer Science and Engineering, IIT Roorkee, India. He has authored or coauthored more than 200 articles in international journals and conferences. His research interests are pattern recognition, bio-signal analysis, EEG-based pattern analysis, and multilingual text recognition. He is an Associate Editor of the IET Image Processing, IET Biometrics, IEICE Transactions on Information and Systems and Springer Nature Computer Science.\\
\end{IEEEbiography}
\end{document}